\documentclass[lettersize,journal]{IEEEtran}
\usepackage{amsmath,amsfonts}
\usepackage{algorithmic}
\usepackage{algorithm}
\usepackage{array}
\usepackage[caption=false,font=normalsize,labelfont=sf,textfont=sf]{subfig}
\usepackage{textcomp}
\usepackage{stfloats}
\usepackage{url}
\usepackage{verbatim}
\usepackage{graphicx}
\usepackage{cite}
\usepackage{hyperref}
\usepackage{xcolor}
\makeatletter

\renewcommand{\maketag@@@}[1]{\hbox{\m@th\normalsize\normalfont#1}}%

\makeatother
\begin{document}

\newcommand{\tabincell}[2]{\begin{tabular}{@{}#1@{}}#2\end{tabular}}
   \newtheorem{Property}{\it Property} 
  
 \newtheorem{Proposition}{\bf Proposition}
\newtheorem{remark}{Remark}
\newenvironment{Proof}{{\indent \it Proof:}}{\hfill $\blacksquare$\par}

\title{Cross-field SNR Analysis and Tensor Channel Estimation for Multi-UAV Near-field Communications}

\author{Tianyu Huo, Jian Xiong, ~\IEEEmembership{Senior Member, IEEE}, Yiyan Wu, ~\IEEEmembership{Life Fellow, IEEE}, Songjie Yang, 

Bo Liu, ~\IEEEmembership{Senior Member, IEEE}, Wenjun Zhang, ~\IEEEmembership{Fellow, IEEE}, 
\thanks{
Tianyu Huo is with the Institute of Image Communication and Network
Engineering, Shanghai Jiao Tong University, Shanghai 200240, China. (e-mail: huotianyu@sjtu.edu.cn).

Jian Xiong is with the Institute of Image Communication and Network
Engineering, Shanghai Jiao Tong University, Shanghai 200240, China; he is also with Zhengzhou Industrial Technology Research Institute of SJTU (e-mail: xjarrow@sjtu.edu.cn).\par

Yiyan Wu is with the Department of Electrical and Computer Engineering, Western University, London, Ontario, Canada. 
 (email: Yiyan.wu@ieee.org)\par
 Songjie Yang is with the National Key Laboratory of Wireless Communications, University of Electronic Science and Technology of China, Chengdu 611731, China. (e-mail: yangsongjie@std.uestc.edu.cn).\par
Bo Liu is with the School of Computer Science, University of Technology
Sydney, Ultimo, NSW 2007, Australia. (e-mail: bo.liu@uts.edu.au).\par
Wenjun Zhang is with the Department of Electronic Engi-
neering, Shanghai Jiao Tong University, Shanghai 200240, China. (e-mail:zhangwenjun@sjtu.edu.cn).

}}

\markboth{Journal of \LaTeX\ Class Files,~Vol.~14, No.~8, August~2021}%
{Shell \MakeLowercase{\textit{et al.}}: SNR analysis and near-field channel estimation for multi-UAV communications}

\IEEEpubid{0000--0000/00\$00.00~\copyright~2021 IEEE}

\maketitle

\begin{abstract}
 Extremely large antenna array (ELAA) is key to enhancing spectral efficiency in 6G networks. Leveraging the distributed nature of multi-unmanned aerial vehicle (UAV) systems enables the formation of distributed ELAA, which often operate in the near-field region with spatial sparsity, rendering the conventional far-field plane wave assumption invalid. This paper investigates channel estimation for distributed near-field multi-UAV communication systems. We first derive closed-form signal-to-noise ratio (SNR) expressions under the plane wave model (PWM), spherical wave model (SWM), and a hybrid spherical–plane wave model (HSPWM), also referred to as the cross-field model, within a distributed uniform planar array (UPA) scenario. The analysis shows that HSPWM achieves a good balance between modeling accuracy and analytical tractability. Based on this, we propose two channel estimation algorithms: the spherical-domain orthogonal matching pursuit (SD-OMP) and the tensor-OMP. The SD-OMP generalizes the polar domain to jointly consider elevation, azimuth, and range. Under the HSPWM, the channel is naturally formulated as a tensor, enabling the use of tensor-OMP. Simulation results demonstrate that tensor-OMP achieves normalized mean square error (NMSE) performance comparable to SD-OMP, while offering reduced computational complexity and improved scalability.

\end{abstract}


\section{Introduction}
\IEEEPARstart{W}{ith} the successful deployment of the fifth-generation (5G) mobile communication systems worldwide, research on the sixth-generation (6G) wireless networks has intensified\cite{6G2}. To support the ambitious objectives of 6G wireless networks\cite{art}\cite{xiao2020overviewintegratedlocalizationcommunication}, such as unprecedented high data rates, ultra-high reliability, global coverage, and ultra-dense connectivity, a variety of new solutions and technologies have recently emerged. Among these, integrated sensing and communications (ISAC)\cite{ISAC}, reconfigurable intelligent surfaces (RIS)\cite{UR}, and extremely large antenna
array (ELAA)\cite{NF}\cite{ELAAdai}, also known as extremely large-scale array (XL-array) \cite{XLarray} are being actively investigated as potential enabling technologies for 6G systems.

Despite their potential to enhance communication performance, these technologies face challenges such as severe free-space path loss and sensitivity to blockage. To achieve ubiquitous 6G coverage, unmanned aerial
vehicles (UAVs) have been widely studied for their mobility, flexibility, and adaptive altitude adjustment \cite{PWM1} \cite{1}. When deployed as aerial base stations (BSs), UAVs enhance wireless coverage and establish favorable propagation conditions, such as line-of-sight (LoS) distribution. However, due to limited payload capacity and LoS-dominated air-to-ground links, a single UAV-enabled antenna array cannot ensure scalable aperture gain and spatial multiplexing \cite{swarm}. Multi-UAV systems address these challenges by extending coverage, ensuring reliable LoS transmission, and improving task efficiency through collaboration while maintaining maneuverability \cite{swarmmm}

Among the numerous envisioned 6G enabling technologies, ELAA has emerged as a promising solution to meet the ultra-high spectral efficiency and spatial resolution demands of future networks \cite{NF}\cite{ELAAtutorial, HOWDOES, elaaa, yang2,yang3}. However, the significant increase in antenna size poses deployment challenges, especially for mobile platforms like vehicles, aircraft, and ships, where space for array antennas is fragmented and limited. To address this, researchers have explored distributed ELAA, including modular ELAA \cite{modular} and widely-spaced multi-subarray (WSMS) concepts \cite{WSMS}\cite{WSMSS}, which enlarge the antenna aperture without additional elements. By dividing a large-aperture array into smaller subarrays deployed across mobile platforms, ELAA performance is maintained while adhering to mobility constraints.

Given this context, we propose to investigate the multi-UAV near-field system, as the increased array aperture brought by multi-UAV makes it easier to enter the near-field, which will present several advantages. First, multi-UAV near-field systems can expand the overall antenna aperture without additional elements, thereby achieving high spatial resolution and enabling precise positioning. Secondly, by equipping each UAV with a medium-sized array, distributed configurations enhance both flight stability and operational flexibility. Finally, this integration complements non-terrestrial networks, fostering the advancement of future wireless communications and their applications. 
\subsection{Related Work}
\subsubsection{Near-field performance analyses}
The deployment of ELAA fundamentally alters channel characteristics, transitioning from far-field to near-field communication, where the conventional plane wave model (PWM) becomes inaccurate. Instead, the spherical wave model (SWM) better captures phase and amplitude variations across large arrays\cite{HOWDOES, elaaa, modular, WSMS, WSMSS}. Prior works have derived closed-form SNR expressions and asymptotic performance analyses under SWM, revealing that SNR scaling deviates significantly from the linear trend predicted by PWM, highlighting the necessity of near-field modeling.

For the distributed ELAA, it is not only capable of accommodating a very large number of antennas, but also meets the practical constraints of the installation structure. More importantly, as the spacing increases, it becomes easier to access the near-field region, which allows for the full exploitation of the advantages inherent to the near-field environment. To accommodate its spatial structure, the hybrid spherical and plane wave model (HSPWM) is adopted, where the planar wavefront is applied to the intra-subarray and the spherical wavefront is used for the inter-subarray\cite{WSMS}\cite{WSMSS}\cite{WS}\cite{WSS}. HSPWM is also referred to as a cross-field model, which captures the transition between near- and far-field regimes. \cite{WSMS} derived the closed-form angle and range Cramér-Rao bounds (CRBs) for the spherical wavefront-based WSMS (SW-WSMS) and HSPWM-based WSMS (HSPWM-WSMS) in bi-static systems. The potential for high-resolution near-field localization using WSMS has been verified through comparative analysis. On the other hand, the application of the HSPWM assumption not only circumvents the inaccuracies associated with the PWM but also mitigates the increased parameterization that arises from the expanded channel matrix dimensions during channel estimation when employing the SWM \cite{WS}\cite{WSS}. Additionally, it also enriches the multiplexing within the propagation paths\cite{WSMSS}. Additionally, by analyzing both near-field and far-field channel models, \cite{crossfield} proposes a two-stage hybrid-field beam training scheme based on coarse beam sweeping and fine hierarchical codebooks. This approach distinguishes near-field users (NUs) from far-field users (FUs) using a modified Rayleigh distance (defined by 90\% beam gain) and 3dB spatial angle width.

\subsubsection{Near field channel estimation}
Accurate Channel State Information (CSI) is vital for ELAA, but conventional far-field estimation methods degrade significantly due to the near-field effect. Recent works have proposed dedicated near-field solutions \cite{yang5,NF1,  NF3}. These include methods based on joint angle-distance sparse representations \cite{NF1} and low-complexity approaches that decouple these parameters for UPAs \cite{NF3}.
 
All of these methods are based on uniform ELAA, where antenna elements are deployed with uniform inter-element spacing, such as in uniform linear or planar arrays.
In contrast, research has also been conducted on channel estimation methods for distributed ELAA, where subarrays are geographically separated.
Literature \cite{WS} divided array antennas into smaller subarrays, treating each subarray’s channel as a far-field channel. A two-stage channel estimation scheme was used, with a deep convolutional neural network estimating subarray channel parameters in the first stage and geometric relationships completing the estimation in the second. To simplify complexity, the HSPWM assumption was employed \cite{WSMS}\cite{yang1}, enabling a two-dimensional orthogonal matching pursuit (2D-OMP) method with lower complexity than traditional polar domain orthogonal matching pursuit (PD-OMP) \cite{yang1}, reducing the computational burden for practical applications.


Near-field channel estimation becomes particularly critical in multi-UAV-assisted communication systems due to the limited altitude, rapid topology changes, and short communication distances, which frequently result in near-field propagation conditions. Several recent studies have applied deep reinforcement learning and generative models to optimize multi-UAV beamforming and secure communications in complex environments \cite{ZhangTMC2025, LiuTMC2024}. These works emphasize the growing importance of robust and adaptive channel estimation techniques under near-field and distributed multi-UAV scenarios.

\subsection{Contributions}
Although existing literature has extensively explored subarray-based channel modeling, the integration of distributed characteristics and near-field propagation conditions in multi-UAV communication systems remains underexplored.  Given that individual UAV is typically equipped with UPA, it is of significant importance to investigate the performance of multi-UAV near-field communication systems. Moreover, existing UAV channel estimation methods are mostly based on far-field assumptions, primarily due to the limited number of antennas on a single UAV and the considerable distance from the ground, which rarely satisfy near-field conditions. 
However, due to the increased aperture caused by UAV deployment scale and spacing, multi-UAV scenarios are likely to operate in the near-field region. For example, in a \(2 \times 2\) UAV configuration with inter-UAV spacing of \(50d\) ($d$ denotes the half-wavelength), the Rayleigh distance exceeds 110~m at 10~GHz. This motivates our research on near-field channel estimation for multi-UAV systems. Consequently, this work establishes the first comprehensive framework for analyzing both performance metrics and channel estimation in multi-UAV near-field systems. To our knowledge, it pioneers the application of ELAA principles to practical UAV swarm configurations. We have made the following contributions:

\begin{itemize}
\item{We model a multi-UAV to user communication framework in which each UAV is equipped with a UPA. The UAV spacing is larger than half-wavelength to cooperatively perceive the user in the ground. To ensure general applicability, we adopt a rectangular formation for the UAV swarm configuration.}
\item{
We investigate the mathematical modeling and SNR performance analysis in multi-UAV near-field communication system. The analytical expressions for the maximum SNR under three models including SWM, PWM and HSPWM are first derived. Performance analysis reveals the unique potential of HSPWM in distributed ELAA systems, characterized by its simplicity and high precision. Notably, this is the first work to conduct performance analysis under the assumption of HSPWM for the distributed ELAA and to compare it comprehensively with both PWM and SWM.}
\item{We propose a channel estimation method applying the HSPWM assumption. First, we extend the PD-OMP method directly to the multi-UAV near-field scenario, thereby developing the spherical domain orthogonal matching pursuit (SD-OMP) method. However, it does not fully exploit the structural characteristics of the near-field array response, leading to a large dictionary size and consequently high complexity in channel estimation. To address this, we further propose a tensor dictionary based on HSPWM, facilitating the development of a low-complexity near-field channel estimation method. Moreover, we employ the Nelder-Mead algorithm to resolve the mismatch issue.
}
\end{itemize}

The paper is organized as follows. Section II introduces the system model of the multi-UAV distributed near-field system. Section III analyzes the SNR performance. Channel estimation methods are proposed in Section IV. Simulation results are shown in Section V. Section VI concludes the paper.

\textbf{Notations:}
\( a \), \( \boldsymbol{a} \), \( \boldsymbol{A} \)  and \( \boldsymbol{\mathcal{A}} \) represent the scalar, the vector, the
matrix and the tensor, respectively. 
 \( \mathbb{C}^{M \times N} \) depicts the set of \( M \times N \)-dimensional complex-valued matrices. 
 \( (\cdot)^{T} \), 
 \( (\cdot)^{H} \) and \( (\cdot)^{{-1}} \)
represent transpose, conjugate
transpose and inverse operation, respectively. \( (\cdot)^{{\dagger}}\) denotes the pseudoinverse, and its formula is $\boldsymbol{A}^{\dagger} = (\boldsymbol{A}^{\mathrm{H}} \boldsymbol{A})^{-1} \boldsymbol{A}^{\mathrm{H}}$. 
 \( \exp\{\cdot\} \) defines the exponential function of \( e \).
 \( \mathbf{I}_N \) defines an \( N \)-dimensional identity matrix.
$\sum_{n=0}^{N}(\cdot) $
 represents the sum of all function values from 0 to $N$.
 \( \circ  \) represents vector outer product.  \( \otimes \) represents Kronecker product. \( \times_1  \), \( \times_2  \) and \( \times_3\) represent the 1-mode, 2-mode and 3-mode product of a tensor with a vector, respectively. 
 \( \|\cdot\| \) and \( \|\cdot\|_F \) stands for the 2-norm and the Frobenius norm. \( \mathcal{CN}(\boldsymbol{a}, \boldsymbol{A}) \) denotes the Gaussian distribution in the complex domain, where $\boldsymbol{a}$ represents the mean, and $\boldsymbol{A}$ signifies the covariance matrix. \( \mathcal{U}\left(-a, a\right) \) signifies a uniform distribution across the interval $(-a,a)$.

\section{SYSTEM MODEL}

As shown in Fig. 1, we consider a near-field SIMO wireless communication system based on uplink time division duplexing (TDD), which comprises  $M=M_x\times M_y$ UAVs and a single-antenna user. Each UAV is equipped with a UPA of $N=N_x\times N_y$ antennas mounted at its bottom, which is shown in Fig. 1. Specifically, the UAV located in the \(m_x\)-th row and \(m_y\)-th column is indexed by \((m_x, m_y)\), \(m_x = 1, 2, \cdots, M_{x}\) and \(m_y = 1, 2, \cdots, M_{y}\). The index of the antenna in the \(n_x\)-th row and \(n_y\)-th column of a UAV is denoted as \((n_x, n_y)\), \(n_x = 1, 2, \cdots, N_x\) and \(n_y = 1, 2, \cdots, N_y\). For simplicity in notation, in the following paper, the UAV positioned at the \(m_x\)-th row and \(m_y\)-th column is referred to as \(m\), where \(m = (m_x-1) M_x + m_y\); each antenna element on the UAV at the \(n_x\)-th row and \(n_y\)-th column is denoted by \(n\), where \(n = (n_x-1)  N_x + n_y\), thereby ensuring the uniqueness of the mapping relationship.  Each UAV carries a radio frequency (RF) chain, with each linked to $\textit{N}$ antenna elements. Therefore, the total number of array elements is $\textit{MN}$. For analytical simplicity, we assume that all UAVs are perfectly synchronized in time and frequency and remain in a stable hovering state with no relative motion. \footnote{All UAVs are assumed to upload their received pilot signals to a central node (e.g., a ground station or a leading UAV) for centralized processing\cite{tongbu}. The centralized scheme assumes that UAVs only upload a small set of sparse channel parameters (e.g., user position and path gains) during predefined time slots, and the central node reconstructs the channel accordingly \cite{ref2}. }

All communication nodes are placed in the three-dimensional (3D) Cartesian coordinate system. The UAV located in the 1-th row and 1-th column, denoted by $\textit{U}_1$, serves as the reference. Note that UAVs should not ascend and descend frequently to avoid energy consumption, so we assume that the UAV swarm flies at a specific altitude of $\textit{H}$. 

For each UAV, the first antenna is chosen as the reference antenna. The antenna spacing $\textit{d}=\lambda/2$ within a UAV, where $\lambda$ denotes the carrier wavelength. The UAV spacing is arbitrarily multiple of half-wavelength, which represents the general setting of the distributed ELAA. Therefore, the distances of the reference antenna element between adjacent UAVs alone on the $x$ axis and the $y$ axis are represented as $D_{x}d$ and $D_{y}d$, respectively, where $D_{x}$, $D_{y} \gg 1$. So the relative position of the $(m_x, m_y)$-th UAV with reference to $\textit{U}_1$ is denoted by $\boldsymbol{d}_{m}=[(m_{x}-1)D_{x}d, (m_{y}-1)D_{y}d, 0]^T$. Denoting the location of $\textit{U}_1$ as $\boldsymbol{p}_{U_1}$, the location of the $(n_x, n_y)$-th antenna on the $(m_x, m_y)$-th UAV is $\boldsymbol{p}_{m, n}=\boldsymbol{p}_{U_1}+\boldsymbol{d}_{m}+\boldsymbol{d}_{n}$,
where $\boldsymbol{d}_{n}=\left[(n_{x}-1)d, (n_{y}-1)d, 0\right]^T$ is the position of the antenna relative to the first antenna of that UAV.
\begin{figure}[!t]
\vspace{-0.3cm} 
\centering
\subfloat[]{\includegraphics[width=0.5\linewidth]{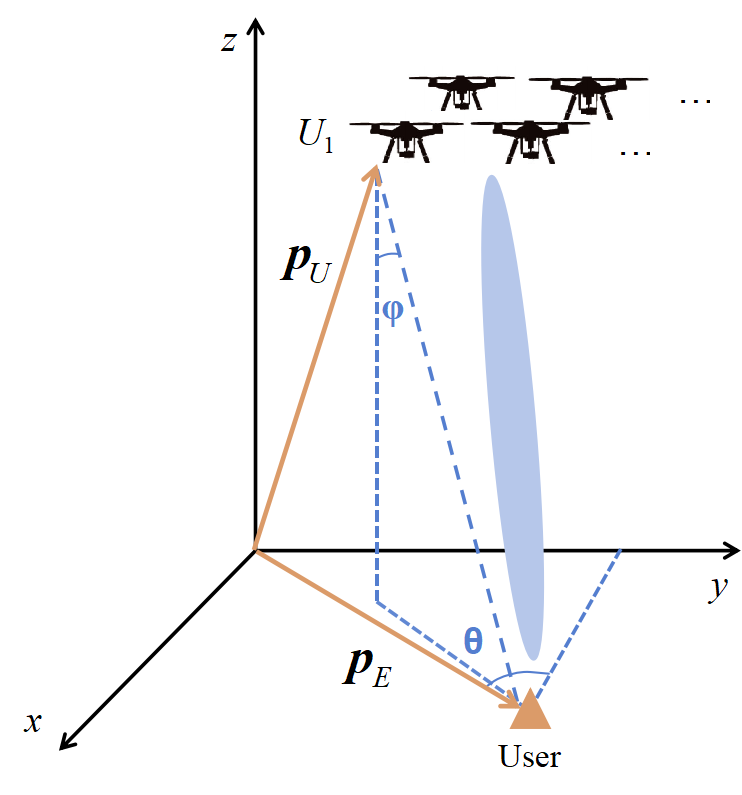}%
\label{fig_first_case}}
\hfil
\subfloat[]{\includegraphics[width=0.5\linewidth]{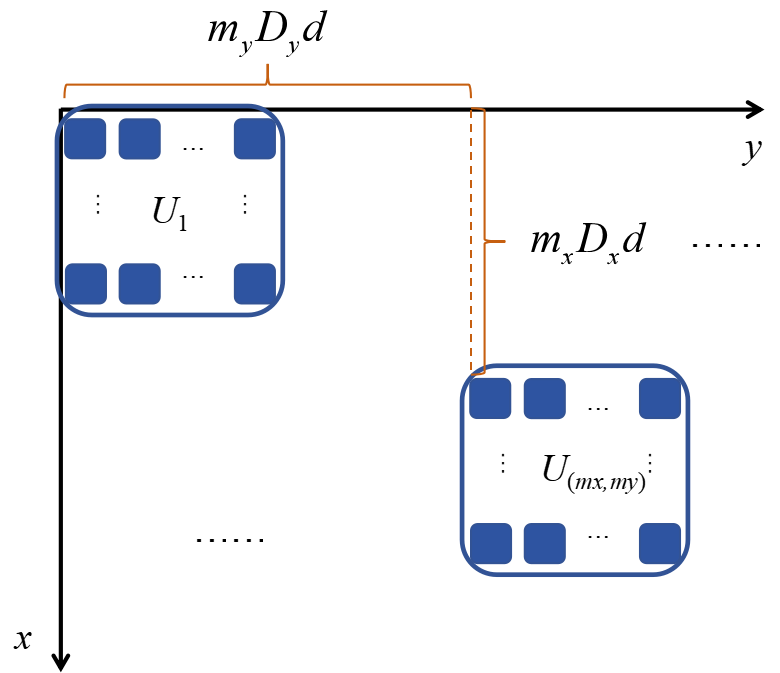}%
\label{fig_second_case}}
\caption{A multi-UAV near-field system, where $\textit{M}$ UAVs cooperatively perceive the target. (a) 3D overview. (b) \emph{xy}-plane overview.}
\label{fig_sim}
\vspace{-0.3cm}
\end{figure}
The user's location is denoted as 
 $\boldsymbol{p}_E$, and its relative position to $\textit{U}_1$ is $\boldsymbol{p}_{U_1E} =  \boldsymbol{p}_{U_1} - \boldsymbol{p}_{E}  =\left[ r \Psi, r \Phi, r \Omega \right]^T$, where $r=\left\| \boldsymbol{p}_{U_1} - \boldsymbol{p}_{E} \right\|$, $\Psi=\sin \varphi \cos \theta$, $\Phi=\sin \varphi \sin \theta$, and $\Omega=\cos \varphi$, with $\theta \in [0, \pi]$ and $\varphi \in \left[-\frac{\pi}{2}, \frac{\pi}{2}\right]$ denote the azimuth and the elevation angle, respectively. Thus the distance between the user and the $(n_x,n_y)$-th antenna element of the $(m_x,m_y)$-th UAV is then given by
 \begin{small}
\begin{equation}
\begin{aligned}
r_{m, n} &= \left\| \boldsymbol{p}_{m, n} - \boldsymbol{p}_{E} \right\| \\
&= r \sqrt{
\begin{split}
1 + 2\xi \Psi (m_x D_x + n_x)  + 2\xi \Phi (m_y D_y + n_y) \\
 + \left[ (m_x D_x + n_x)^2 + (m_y D_y + n_y)^2 \right] \xi^2,
\end{split}
}
\end{aligned}
\vspace{-1mm}
\end{equation}
\end{small}
where $\xi = \frac{d}{r}\ll 1$, and $r=r_{1,1}$  is the distance between the user and the reference antenna of $\textit{U}_1$.

\subsection{Signal model}

Assuming that the pilot signal sent by the user at  \( q \)-th time slot is denoted as \( x_q \), the received signal \( y_{q,m} \) on the \( m \)-th UAV can be expressed as:
\begin{small}
\begin{equation}
y_{q,m} = \boldsymbol{w}_{q,m}^H \boldsymbol{h}_m x_q + \boldsymbol{w}_{q,m}^H \boldsymbol{n}_{q,m},
\end{equation}
\end{small}
where \( \boldsymbol{w}_{q,m}\in \mathbb{C}^{N \times 1} \) represents the measurement combiner at the UAV side , and $\boldsymbol{h}_{m}  \in \mathbb{C}^{N\times1} $ denotes the wireless channel between the $m$-UAV and the user. \( \boldsymbol{n}_{q,m} \) denotes the complex additive white Gaussian noise (AWGN), which follows the distribution \( \mathcal{CN}(0,\sigma^2\mathbf{I}_N) \).

Assuming that the pilot signal \( {x}_q \) sent by the user is 1 and  \( Q \) is denoted as the pilot length, the collection of all received signals \( \boldsymbol{y}_m \) on the \( m \)-th UAV can be expressed as:
\begin{small}
\begin{equation}
\setlength{\abovedisplayskip}{0.2pt}
\setlength{\belowdisplayskip}{0.2pt}
\boldsymbol{y}_{m} = \boldsymbol{W}_{m}^H \boldsymbol{h}_{m} + \overline{\boldsymbol{n}}_{m},
\end{equation}
\end{small}
where $\boldsymbol{W}_m = [\boldsymbol{w}_{1,m}, \ldots, \boldsymbol{w}_{Q,m}] \in \mathbb{C}^{N \times Q}$ denotes the measurement matrix of the $m$-th UAV subarray.  For a UPA with $N=N_x N_y$, 
the measurement matrix $\boldsymbol{W}_m$ is designed as $\boldsymbol{W}_m=\boldsymbol{W}_x \otimes \boldsymbol{W}_y$, with $\boldsymbol{W}_x \in\mathbb{C}^{N_x \times Q_x} $ and $\boldsymbol{W}_y \in\mathbb{C}^{N_y \times Q_y} $ representing the measurement matrix along the $x$-axis and $y$-axis, respectively. $Q_x$ and $Q_y$ denote the number of measurements along the $x$-axis and $y$-axis, respectively. Additionally, $\overline{\boldsymbol{n}}_m = [\boldsymbol{w}_{1,m}^H \boldsymbol{n}_{1,m}, \ldots, \boldsymbol{w}_{Q,m}^H \boldsymbol{n}_{Q,m}]^T \in \mathbb{C}^{Q \times 1}$ represents the noise projection.

According to compressed sensing theory, to enable efficient channel estimation, the measurement matrices \( \boldsymbol{W}_x \) and \( \boldsymbol{W}_y \) are required to exhibit low mutual coherence. Taking \( \boldsymbol{W}_x \) as an example (the construction of \( \boldsymbol{W}_y \) follows similarly), we construct \( \boldsymbol{W}_x \in \mathbb{C}^{N_x \times Q_x} \) based on singular value decomposition (SVD) as:
 \begin{small}
\begin{equation}\label{W1}
\setlength{\abovedisplayskip}{-1.5pt} 
\boldsymbol{W}_x = \boldsymbol{U}_x \boldsymbol{\Sigma}_x \boldsymbol{V}_x^H,
\vspace{-0.25cm}
\end{equation}
 \end{small}
where \( \boldsymbol{U}_x \in \mathbb{C}^{N_x \times N_x} \) and \( \boldsymbol{V}_x \in \mathbb{C}^{Q_x \times Q_x} \) are unitary matrices, and the singular value matrix is given by
$
\boldsymbol{\Sigma}_x = \begin{bmatrix}
\mathrm{diag}(\boldsymbol{\sigma}) ,
\mathbf{0}_{Q_x,\, N_x - Q_x}
\end{bmatrix}^T $, and$\quad \boldsymbol{\sigma} = [\sigma_1, \ldots, \sigma_{Q_x}]^T$ and $\mathbf{0}_{Q_x,\, N_x - Q_x}\in \mathbb{C}^{Q_x \times (N_x-Q_x)}$ is a null matrix.

To ensure orthogonality,  the following Frobenius norm metric is introduced:
 \begin{small}
\begin{equation}\label{W2}
\setlength{\abovedisplayskip}{-1.5pt} 
\left\| \boldsymbol{I}_{Q_x} - \boldsymbol{W}_x^H \boldsymbol{W}_x \right\|_F^2
= \left\| \boldsymbol{I}_{Q_x} - \boldsymbol{\Sigma}_x \boldsymbol{\Sigma}_x^H \right\|_F^2
= \sum_{i=1}^{Q_x} (1 - \sigma_i^2)^2. 
\end{equation}
 \end{small}
This expression is minimized when all singular values equal one, i.e., \( \sigma_1 = \cdots = \sigma_{Q_x} = 1 \). Accordingly, the unconstrained optimal solution is given by:
 \begin{small}
\begin{equation}\label{W3}
\boldsymbol{W}_x = \boldsymbol{U}_x \begin{bmatrix}
\boldsymbol{I}_{Q_x} ,
\mathbf{0}_{Q_x,\, N_x - Q_x}
\end{bmatrix}^T \boldsymbol{V}_x^H,
\end{equation}
 \end{small}
where \( \boldsymbol{U}_x \in \mathbb{C}^{N_x \times N_x} \) and \( \boldsymbol{V}_x \in \mathbb{C}^{Q_x \times Q_x} \) are arbitrary unitary matrices.

To meet the modulus-1 constraint imposed by phased-array hardware, each element of \( \boldsymbol{W}_x \) is normalized as:
 \begin{small}
\begin{equation}\label{W4}
[\boldsymbol{W}_x]_{n_x,q_x} \leftarrow \frac{[\boldsymbol{W}_x]_{n_x,q_x}}{|[\boldsymbol{W}_x]_{n_x,q_x}|}.
\vspace{-0.1cm}
\end{equation}
 \end{small}

By collecting signals from all UAVs, the received signal $\boldsymbol{y}$ can be represented as:
 \begin{small}
\begin{equation}\label{RS}
\setlength{\abovedisplayskip}{-1.5pt} 
\boldsymbol{y} = \boldsymbol{W}^H \boldsymbol{h} + \overline{\boldsymbol{n}},
\end{equation}
 \end{small}
where $\boldsymbol{y} = \begin{bmatrix} \boldsymbol{y}_1^T, \ldots, \boldsymbol{y}_M^T \end{bmatrix}^T \in \mathbb{C}^{MQ \times 1}$, $\boldsymbol{h} = \begin{bmatrix} \boldsymbol{h}_1^T, \ldots, \boldsymbol{h}_M^T \end{bmatrix}^T \in \mathbb{C}^{MN \times 1}$ and $\overline{\boldsymbol{n}} = \begin{bmatrix} \overline{\boldsymbol{n}}_1^T, \ldots, \overline{\boldsymbol{n}}_M^T \end{bmatrix} \in \mathbb{C}^{MQ \times 1}$. Here, $\boldsymbol{W}\in \mathbb{C}^{MN \times MQ}$ denotes the overall measurement matrix composed of 
$M$ subarrays. In this study, we assume that each subarray employs the same measurement matrix, i.e., $\boldsymbol{W}_0=\boldsymbol{W}_1=\ldots=\boldsymbol{W}_{M-1}$. Consequently, $\boldsymbol{W}$  has a block diagonal structure and can be streamlined to $\boldsymbol{W}=\mathbf{I}_{M} \otimes \boldsymbol{W}_x \otimes \boldsymbol{W}_y$, which can be expressed as 
 \begin{small}
\begin{equation}\label{W6}
\boldsymbol{W} = \begin{bmatrix}
\boldsymbol{W}_0 & 0 & \cdots & 0 \\
0 & \boldsymbol{W}_1 & \cdots & 0 \\
\vdots & \vdots & \ddots & \vdots \\
0 & 0 & \cdots & \boldsymbol{W}_{M-1}
\end{bmatrix}.
\vspace{-0.35cm}
\end{equation}
 \end{small}

\subsection{Channel model}
Considering the geometric structure and the nature of limited scattering, the channel $\boldsymbol{h}$ is written as
 \begin{small}
\begin{equation}\label{CM}
\setlength{\abovedisplayskip}{0.5pt}
\setlength{\belowdisplayskip}{0.2pt}
\boldsymbol{h} = \frac{1}{\sqrt{L}} \sum_{l=1}^{L} z_{l} \boldsymbol{g}(\theta_{l},\varphi_l, r_{l}),
\end{equation}
 \end{small}
where $\textit{L}$ is the number of channel paths, $\textit{z}_l$ denote the complex gain of the $\textit{l}$-th path. As depicted in Fig. 1, \( \{\theta_l\}_{l=1}^L \), \( \{\varphi_l\}_{l=1}^L \) and \( \{r_l\}_{l=1}^L \) denote the azimuth angle, the elevation angle and distance parameters, respectively. The array response $\boldsymbol{g}(\theta, \varphi, r)$ is given by $\boldsymbol{g}(\theta, \varphi, r)=[g_{0,0}(\theta, \varphi, r),\ldots,g_{m,n}(\theta, \varphi, r),\ldots,g_{M,N}(\theta, \varphi, r)]^T\in \mathbb{C}^{MN \times 1}$, which $g_{m, n}\left(\theta, \varphi, r\right)$ is the array response element between the user and the $(n_x, n_y)$-th element antenna in the $(m_x, m_y)$-th UAV. $g_{m, n}\left(\theta, \varphi, r\right)$ can be expressed as
$
{g}_{m, n}(\theta, \varphi, r) = \frac{\sqrt{\beta_{0}}}{r_{m, n}} e^{-j \frac{2\pi}{\lambda} r_{m, n}}
$
with $\beta_{0}$ denoting the
channel power at the reference distance $\textit{d}_0$ = 1 m. 

In this subsection, various forms of channel responses tailored to the unique signal propagation characteristics of different communication scenarios are presented, namely $\boldsymbol{g}_{\mathbf{SWM}}(\theta, \varphi, r)$, $\boldsymbol{g}_{\mathbf{PWM}}(\theta, \varphi, r)$ and $\boldsymbol{g}_{\mathbf{HPSW}}(\theta, \varphi, r)$.

\subsubsection{SWM}
For users in the near-field region, the channel modeling should follow the SWM, which is the most accurate model that calculates the channel response
from all antennas to the user. Consequently, the array response vector $\boldsymbol{g}_{\mathbf{SWM}}\left(\theta, \varphi, r\right) \in \mathbb{C}^{M N \times 1}$ between the user and the distributed ELAA can be given by (\ref{A}), in which, the term $\frac{r}{r_{m, n}}$ determines the magnitude, and the term $r_{m,n}-r$ determines the phase.
\begin{figure*}[ht]
\normalsize
\vspace{-0.5cm}
\centering
 \begin{small}
\begin{equation}\label{A}
\setlength{\abovedisplayskip}{-0.3pt}
\setlength{\belowdisplayskip}{-0.3pt}
\boldsymbol{g}_{\mathbf{SWM}}(\theta, \varphi, r) = \frac{\sqrt{\beta_{0}}}{r} e^{-j\frac{2\pi}{\lambda} r} \left[1,  \ldots,  \frac{r}{r_{m, n}} e^{-j \frac{2\pi}{\lambda} (r_{m, n}-r)},  \ldots, \frac{r}{r_{M-1, N-1}} e^{-j\frac{2\pi}{\lambda}(r_{M-1, N-1} - r)}\right]^T
\end{equation}
 \end{small}
\vspace{-0.35cm}
\end{figure*}
\subsubsection{PWM}
For far-field scenarios where the user-to-array distance surpasses the Rayleigh distance, the channel model is well approximated by the PWM, given the wavefront's near-planar nature.
 Therefore, the amplitude variation caused by $r_{m,n}$ is assumed to be relatively small. In line with most papers investigating far-field scenarios\cite{PWM1}\cite{PWM2}\cite{pwm3}, $\frac{r}{r_{m, n}}$ is set to 1 in this paper. Additionally, since $r_{m,n}-r$ determines the phase, even a subtle change will have great impact, which can be represented as
 \begin{small}
\begin{equation}
\setlength{\abovedisplayskip}{-1.5pt} 
r_{m,n} - r = \left(n_x + m_x D_x\right) d\Psi + \left(n_y + m_y D_y\right) d\Phi.
\vspace{-0.1cm}
\end{equation}
\end{small}
Consequently, the array response vector $\boldsymbol{g}_{\mathbf{PWM}}\left(\theta, \varphi, r\right) \in \mathbb{C}^{M N \times 1}$ for PWM can be expressed as (\ref{B}). 

It is important to recognize that, unlike the SWM, which depends on both amplitude and phase variations, the PWM is solely governed by phase variation. As a result, the PWM serves as an approximation of the SWM when the array size is much smaller than the communication distance. In other words, when the distance between the user and the array is greater than the Rayleigh distance, the SWM degenerates into the PWM. 

\subsubsection{HSPWM}

Drawing from above analyses, while the SWM provides the most accurate depiction of channel characteristics, this detailed modeling introduces significant complexities in channel estimation, posing many challenges. On the other hand, the PWM, despite its analytical simplicity, incurs a loss in precision, which diminishes its applicability in near-field conditions. Recognizing these limitations, researchers have turned their focus to the HSPWM assumption, which offers a more balanced approach \cite{WSMS}. This assumption is gaining traction for its potential to mitigate the challenges associated with both the SWM and PWM, providing a more viable framework for channel estimation.

\begin{figure*}[ht]
\normalsize
\centering
 \begin{small}
\begin{equation}\label{B}
\setlength{\abovedisplayskip}{-0.3pt}
\setlength{\belowdisplayskip}{-0.3pt}
\begin{aligned}
\boldsymbol{g}_{\mathbf{PWM}}(\theta, \varphi, r) =& \frac{\sqrt{\beta_{0}}}{r} e^{-j\frac{2\pi}{\lambda} r} \left[1,  \ldots, e^{-j\frac{2\pi}{\lambda}\left[\left(n_x + m_x D_x\right) d\Psi + \left(n_y + m_y D_y\right) d\Phi\right]},  \ldots,  e^{-j\frac{2\pi}{\lambda}\left[\left(N_{x}-1+\left(M_{x}-1\right) D_{x}\right) d\Psi+\left(N_{y}-1+\left(M_{y}-1\right) D_{y}\right) d\Phi\right]}\right]^T
\end{aligned}
\end{equation}
 \end{small}
\vspace*{-0.5cm}
\end{figure*}

Specifically, under the HSPWM assumption, the PWM is utilized within each subarray, maintaining accuracy due to the small array size. The SWM is then applied to enhance the modeling precision within these subarrays. In our scenario, PWM is applied to the UPA within one UAV, while SWM is utilized among UAVs. Considering the HSPWM assumption, $\boldsymbol{g}_{\mathbf{HSPWM}}(\theta, \varphi, r)\in \mathbb{C}^{M N \times 1}$ is obtained by approximating $\boldsymbol{g}_{\mathbf{SWM}}(\theta, \varphi, r)$ 
\begin{small}
\begin{equation} \label{C}
\begin{aligned}
\boldsymbol{g}_{\mathbf{HSPWM}}(\theta, \varphi, r) &\approx \boldsymbol{b}(\theta, \varphi, r) \otimes \boldsymbol{a}(\theta, \varphi),
\end{aligned}
\end{equation}
\end{small}
where $\boldsymbol{a}(\theta, \varphi) \in \mathbb{C}^{N \times 1}$ denotes the plane-wave array manifold for each UAV, and 
$\boldsymbol{b}(\theta,\varphi, r) \in \mathbb{C}^{M \times 1}$ denotes the spherical-wave array manifold for the inter-UAV. Their expressions are as follows:
\begin{small}
\begin{equation} \label{afj}
\begin{aligned}
\boldsymbol{a}(\theta, \varphi) = \sqrt{\beta_{0}} \left[1, \ldots, e^{-j \frac{2\pi}{\lambda} d(n_x \sin\varphi \sin\theta  + n_y \sin\varphi \cos\theta )}, \ldots, \right.\\[-2mm]
\left. e^{-j \frac{2\pi}{\lambda} d ((N_x-1) \sin\varphi \sin\theta+(N_y-1) \sin\varphi \cos\theta)}\right]^{T},
\end{aligned}
\vspace{-0.3cm}
\end{equation}
\end{small}
\begin{small}
\begin{equation}\label{afjj}
\begin{aligned}
\boldsymbol{b}(\theta, \varphi, r) = &\frac{1}{r_{m,0}} e^{-j\frac{2\pi}{\lambda} r}\left[ 1, \ldots, e^{-j \frac{2\pi}{\lambda} (r_{m,0} - r)},\right.\\[-2mm]
&\left. \ldots, e^{-j \frac{2\pi}{\lambda} (r_{M-1,0} - r)} \right]^T.
\end{aligned}
\vspace{-0.3cm}
\end{equation}
\end{small}
\section{performance analysis of SNR}
In this section, we initially provide a closed-form expression solution for the SNRs of three models, i.e., SWM, PWM and HSPWM. Furthermore, an asymptotic performance analysis is provided to describe the impact of the distribution of UAV array on SNR in our multi-UAV mmwave communication system. Upon comparison, it is evident that the HSPWM delivers the low complexity akin to the PWM, while also retaining the high accuracy of the SWM, thereby conferring significant advantages. 

\subsection{Closed-form expression of SNR}
For downlink communication, considering the free-space LoS propagation, the user's received signal can be expressed as
\begin{small}
\begin{equation}
\setlength{\abovedisplayskip}{-2.5pt}
y = \boldsymbol{g}(\theta, \varphi, r) \sqrt{P} \boldsymbol{f} s + \boldsymbol{\eta}\boldsymbol{f}, 
\vspace{-0.3cm}
\end{equation}
\end{small}
where \( P \) and \( s \) are the transmit power and information-bearing signal of the user, respectively; \( \boldsymbol{\eta} \sim \mathcal{CN}(0, \sigma^2 \mathbf{I}_{MN}) \) is the AWGN with covariance matrix $\sigma^2 \mathbf{I}_{MN}$; \( \boldsymbol{f} \in \mathbb{C}^{MN \times 1} \) is the transmit beamforming vector, with \( \|\boldsymbol{f}\| = 1 \). Note that, although the uplink system model in Sections II, IV, and V employs a single-RF-chain analog combiner at the UAV receiver, we adopt a downlink model with digital beamforming here to facilitate tractable analysis. This allows us to derive closed-form SNR expressions and evaluate the upper-bound performance under ideal beamforming strategies.
Thus the resulting received SNR can be given by $\gamma = P \left|  \boldsymbol{g}(\theta, \varphi, r) \boldsymbol{f} \right|^2/\sigma^2.$

According to the maximal-ratio combining (MRC) strategy\cite{modular}\cite{Xiong},  i.e.,
$
\boldsymbol{f}^* = \frac{\boldsymbol{g}(\theta, \varphi, r)}{\|\boldsymbol{g}(\theta, \varphi, r)\|},
$
the received SNR will reach its maximum value, which is 
\begin{small}
\begin{equation} \label{gamma}
\setlength{\abovedisplayskip}{1pt}
\setlength{\belowdisplayskip}{1pt}
\gamma = \bar{P} \|\boldsymbol{g}(\theta, \varphi, r)\|^2,
\end{equation}
\end{small}
where \( \bar{P} = \frac{P}{\sigma^2} \) is the transmit SNR. 
\begin{figure*}[hb]
\normalsize
\vspace{-0.3cm}
\centering
\begin{small}
\begin{equation}\label{E}
\setlength{\abovedisplayskip}{-1.1pt}
\setlength{\belowdisplayskip}{-0.1pt}
\begin{aligned}
\gamma_{\mathbf{SWM}} =  \frac{\bar{P} \beta_{0}}{r^2} \sum_{m_x=0}^{M_x-1} \sum_{m_y=0}^{M_y-1} \sum_{n_x=0}^{N_x-1} \sum_{n_y=0}^{N_y-1} \frac{1}{1 + 2\xi \Psi (m_x D_x + n_x) + 2\xi \Phi (m_y D_y + n_y)  + \left[ (m_x D_x+ n_x)^2 + (m_y D_y + n_y)^2 \right] \xi^2}
\end{aligned}
\end{equation}
\end{small}
\vspace{-1cm}
\end{figure*}

Following from (\ref{A}) and (\ref{gamma}), the SNR expression for SWM can be given by (\ref{E}). 

\textbf{Proposition 1.}
The expression of the SNR for SWM can be obtained by
\begin{small}
\begin{equation} \label{2int}
\setlength{\abovedisplayskip}{0.2pt}
\setlength{\belowdisplayskip}{0.1pt}
\begin{aligned}
   \gamma_{\mathbf{SWM}} \approx  & \frac{\bar{P} \beta_{0}r^2}{D_x d^4}  \int_{-\frac{\xi}{2} }^{(M_y-\frac{1}{2}) \xi} \int_{-\frac{\xi}{2} }^{(N_y-\frac{1}{2}) \xi} \left[ h \left( \frac{S}{\alpha r}+\frac{\Psi}{\alpha} \right) \right.  \\
     & \left.   
     - h \left( \frac{S  - M_x D_x d}{\alpha r} + \frac{\Psi}{\alpha} \right) 
     \right.  \\
     & \left.   
     - h \left( \frac{S  - N_x d}{\alpha r} + \frac{\Psi}{\alpha} \right) 
     \right.  \\
     &    \left.
     + h \left( \frac{S-N_x d-M_x D_x d}{\alpha r}+\frac{\Psi}{\alpha} \right) \right] dt dy,  
\end{aligned}
\end{equation}
\end{small}
where
\begin{small}
\begin{equation}
\setlength{\abovedisplayskip}{0.1pt}
\setlength{\belowdisplayskip}{0.1pt}
\alpha = \sqrt{1 + 2\Phi D_y y + 2\Phi t + D_y^2 y^2 + 2D_y y t + t^2 - \Psi^2} ,
\vspace{-0.2cm}
\end{equation}
\end{small}
\begin{small}
\begin{equation}
\setlength{\abovedisplayskip}{-0.7pt}
\setlength{\belowdisplayskip}{-0.5pt}
h(\varrho) = \varrho \arctan \varrho - \frac{1}{2} \ln(1 + \varrho^2),
\vspace{-0.5cm}
\end{equation}
\end{small}
\begin{small}
\begin{equation}
\setlength{\abovedisplayskip}{-0.5cm}
\setlength{\belowdisplayskip}{-0.5cm}
A_{px} = (M_x - 1) D_x d + (N_x - 1) d,
\vspace{-0.2cm}
\end{equation}
\end{small}
and $S=A_{px}+\frac{D_x d}{2}+\frac{d}{2}$, which $A_{px}$ represents the array aperture alone the $x$-axis, and $\varrho$ is the argument of the function 
$h(\cdot)$.

Proof: Please refer to Appendix A of the supplemental material.

Following from (\ref{B}) and (\ref{gamma}), the SNR expression for PWM can be calculated as 
\begin{small}
\begin{equation} \label{pwm}
\setlength{\abovedisplayskip}{0.2pt}
\setlength{\belowdisplayskip}{0.3pt}
\gamma_{\mathbf{PWM}}  = \frac{\bar{P} \beta_0 MN}{r^2 }.
\end{equation}
\end{small}

Incorporating (\ref{C}), (\ref{afj}) and (\ref{afjj}) into (\ref{gamma}), the SNR expression for HSPWM can be given by (\ref{F}).
\begin{figure*}[ht]
\centering
\begin{small}
\begin{equation}\label{F}
\begin{aligned}
\gamma_{\mathbf{HSPWM}} = \bar{P} \| \boldsymbol{b}(\theta, \varphi, r) \otimes \boldsymbol{a}(\theta, \varphi) \|^2 = \frac{\bar{P} \beta_0 N}{r^2 } \sum_{m_x=0}^{M_x-1} \sum_{m_y=0}^{M_y-1} \frac{1}{  1 + 2\xi \Psi m_x D_x + 2\xi \Phi m_y D_y + \left( (m_x D_x)^2 + (m_y D_y)^2 \right) \xi^2 }
\end{aligned}
\end{equation}
\end{small}
\vspace{-0.3cm}
{\rule{\linewidth}{0.4pt}}
\end{figure*}

To obtain a clearer analytical solution, the scenario is simplified by setting $M_y=1$ and 
$N_y=1$. Consequently, 
$M$ UAVs are arranged in a row alone the $x$-axis, each carrying a ULA with $N=N_x \times 1$ elements. Under these conditions, we derive the analytical SNR expression for three models.

\textbf{Proposition 2.}
 The analytical SNR expression for SWM under the case where the UAV carries a ULA can be derived as 
 \begin{small}
\begin{equation}\label{H}
\setlength{\abovedisplayskip}{0.1pt}
\begin{aligned}
\gamma_{\mathbf{SWM,ULA}} \approx  &\frac{\bar{P} \beta_{0}}{D_x d^2}  \left[ h \left( \frac{S }{r \sqrt{1 - \Psi^2}} + \frac{\Psi}{\sqrt{1 - \Psi^2}} \right)\right.\\ &
\left.  - h \left( \frac{S - M_x  D_x d}{r \sqrt{1 - \Psi^2}} + \frac{\Psi}{\sqrt{1 - \Psi^2}} \right) \right.\\ &
\left.
 - h \left( \frac{S  - N_x  d}{r \sqrt{1 - \Psi^2}} + \frac{\Psi}{\sqrt{1 - \Psi^2}} \right) \right.\\ &
\left.
 + h \left( \frac{S  - N_x  d- M_x  D_x d}{r\sqrt{1 - \Psi^2}} +\frac{\Psi}{\sqrt{1 - \Psi^2}}\right) \right].
\end{aligned}
\end{equation}
\end{small}

Proof: Please refer to Appendix B of the supplemental material.

\emph{Remark: It is worth noting that (\ref{H}) is structurally similar to (\ref{2int}). The difference lies in the fact that the ULA is less significantly affected by the distance $r$ in the context of SWM compared to $\gamma_{\mathbf{SWM}}$ in (\ref{2int}). From (\ref{H}), it can be observed that the maximum SNR for multi-UAV distributed near-field communication is determined by the geometric configuration of the array, such as the total array size \( S \) and the UAV spacing \( D_x d \), as well as the distance \( r \) between the user and $U_1$, the azimuth angle $\theta$ and elevation angle $\varphi$.}

The SNR expression for PWM when the UAV is equipped with a ULA remains consistent with  (\ref{pwm}), i.e., $\gamma_{\mathbf{PWM,ULA}}  = \frac{\bar{P} \beta_0 M_xN_x}{r^2 }$,   which increases linearly with $MN$ and becomes unbounded as $MN $ approaches infinity. This result is clearly impractical, fundamentally attributable to the inapplicability of the PWM assumption when $MN$ is large, leading the communication scenario into the near-field domain. Additionally, it is crucial to recognize that while $\gamma_{\mathbf{SWM,ULA}}$ depends on both the distance $r$ and the geometric configuration of the array, $\gamma_{\mathbf{PWM,ULA}}$ is solely dependent on the distance $r$.

\textbf{Proposition 3.}
The analytical SNR expression for HSPWM can be calculated as
\begin{small}
\begin{equation} \label{Iii}
\setlength{\abovedisplayskip}{-0.1pt}
\begin{aligned}
      \gamma_{\mathbf{HSPWM,ULA}} \approx  & \frac{\bar{P} \beta_{0}N}{r D_x d \sqrt{1 - \Psi^2}} \left[ \arctan \left( \frac{ S_D +\frac{1}{2} D_xd }{r \sqrt{1 - \Psi^2}} \right. \right. \\[-1mm]
     & \left.   \left.
      + \frac{\Psi}{\sqrt{1 - \Psi^2}} \right) \right. \\[-1mm]
      & \left.  
      - \arctan \left( \frac{ S_D -(M_x-\frac{1}{2}) D_x d }{r \sqrt{1 - \Psi^2}}\right. \right. \\[-1mm]
     & \left.   \left.+
      \frac{\Psi}{\sqrt{1 - \Psi^2}} \right) \right], 
\end{aligned}
\end{equation}
\end{small}
where $S_D=(M_x-1)D_x d$.

Proof: Please refer to Appendix C of the supplemental material.

\emph{Remark: The $S_D$ can be understood as the aperture of a $M_x \times 1$ ULA with a spacing of $D_xd$. Comparing (\ref{H}) and (\ref{Iii}), it is evident that, different from SNR which is based on the SWM, the SNR based on the HSPWM assumption is solely related to the aperture $S_D$, and the impact of $N$ on SNR is linear. In other words, by jointly comparing (\ref{pwm}), (\ref{H}) and (\ref{Iii}), the HPSWM integrates the characteristics of both the PWM and the SWM. Not only does it exhibit a linear increase with the growth of $N$, but also depends on the geometric configuration of the ELAA, such as the aperture $S_D$, the UAV spacing $D_xd$, as well as the distance \( r \) between the user and $U_1$, the elevation and azimuth angles.}

\subsection{Performance analysis}
To further investigate the superiority of HSPWM in the multi-UAV near-field system, numerical simulations are developed to evaluate the analytical results derived in the previous section.

To enhance the generality of the scenario, we have configured the UAVs with a 4$\times$4 structure, i.e., $M_x=M_y=4$. The number of antennas for each UAV is set to $N_x=N_y=12$. The carrier frequency is 10 GHz, and the inter-antenna spacing within the UAV is $d=\frac{\lambda}{2}=0.015$ m. The spacing between the UAVs alone the $x$ and $y$ directions is $D_x d=D_y d=50d=0.75$ m, and the coordinates of the user is set as $\boldsymbol{p}_E=[0, 0, 0]^T$.   

Fig. 2 illustrates the SNRs $\gamma_{\mathbf{PWM}}, \gamma_{\mathbf{SWM}}$ and $\gamma_{\mathbf{HPSWM}}$ versus the UAV spacing $D_x d$, with different angles $\varphi$ and $\theta$. As shown in Fig. 2, $\gamma_{\mathbf{PWM}}$ is independent of $D$, $\varphi$ and $\theta$, which is in accordance with the theoretical analysis. In contrast, the SNRs $\gamma_{\mathbf{SWM}}$ and $ \gamma_{\mathbf{HPSWM}}$ exhibit different trends for various UAV orientations $\varphi$ and $\theta$. It is also observed that $\gamma_{\mathbf{PWM}}$ surpasses $\gamma_{\mathbf{SWM}}$ when $\varphi = -30^{\circ}$ and $\theta = 0^{\circ}$, but falls below $\gamma_{\mathbf{SWM}}$ when $\varphi = 45^{\circ}$ and $\theta = 30^{\circ}$. This demonstrates the necessity for near-field analysis in ELAA communication. Additionally, it is noted that $\gamma_{\mathbf{HSPWM}}$ closely aligns with $\gamma_{\mathbf{SWM}}$ regardless of the spacing variation. According to the previous theoretical analysis, the
HSPWM exhibits the lower theoretical complexity compared to
the SWM and better reflects the SNR trend in the near field,
offering significant advantages. Furthermore, as observed in Fig. 2 and indicated by (\ref{H}), the $\gamma_{\mathbf{SWM}}$ can be either higher or lower than its PWM counterpart, depending on the directional cosine $\Psi$, which reflects the angular sensitivity of SWM.

Fig. 3 illustrates the SNRs $\gamma_{\mathbf{PWM}}, \gamma_{\mathbf{SWM}}$ and $\gamma_{\mathbf{HPSWM}}$ versus the element number $M N$ by increasing the number of UAVs $M$ with different angles  $\varphi$ and $\theta$. As shown in Fig. 3, under the same conditions $\varphi$ and $\theta$, for the relatively small element number $MN$, the three models increase with the same growth trend. However, as $MN$ further increases, the SNR based on the PWM and the SNRs based on SWM and HSPWM exhibit two different growth trends. 
Furthermore, it is observed that the SNRs based on the SWM and HSPWM models show different trends for different UAV orientations $\varphi$ and $\theta$, while $\gamma_{\mathbf{PWM}}$ is independent of $\varphi$ and $\theta$.
 Importantly, regardless of the element number variation, $\gamma_{\mathbf{HPSWM}}$ coincides exactly with $\gamma_{\mathbf{SWM}}$. Both Figs. 2 and 3 show that as the UAV spacing and the number of antenna elements change, the SNR gap between the PWM and the SWM/HSPWM increases, which confirms the necessity of near-field analysis in distributed ELAA. It is also worth noting that the HSPWM and SWM curves always coincide under practical UAV configurations\footnote{The overlap of HSPWM and SWM curves results from the limited number of antennas per UAV, which ensures the far-field assumption within each subarray remains valid in practice.
}. This is because the number of antennas mounted on a single UAV is physically limited (e.g., a typical UAV frame size of 47 cm × 58 . 5 cm × 21 . 5 cm (e.g., based on DJI Matrice 30) only supports up to around 31 elements), making the planar-wave assumption within each subarray in HSPWM sufficiently accurate. Therefore, HSPWM not only maintains modeling precision but also offers reduced complexity, validating its suitability as an efficient substitute for SWM in practical distributed ELAA scenarios.

 \begin{figure}[!t]
 \vspace{-0.3cm} 
\centering
\includegraphics[width=0.8\linewidth]{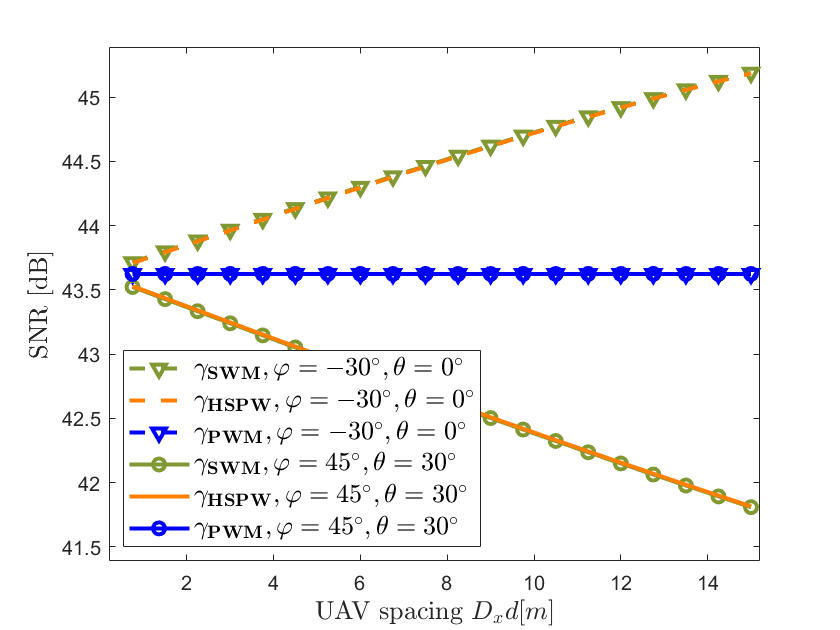}
\caption{ SNRs versus the UAV spacing $D_x d$ with different models and different $\varphi$ and $\theta$.}
\label{fig_2}
\vspace{-0.5cm}
\end{figure}

\begin{figure}[!t]
\centering
\includegraphics[width=0.8\linewidth]{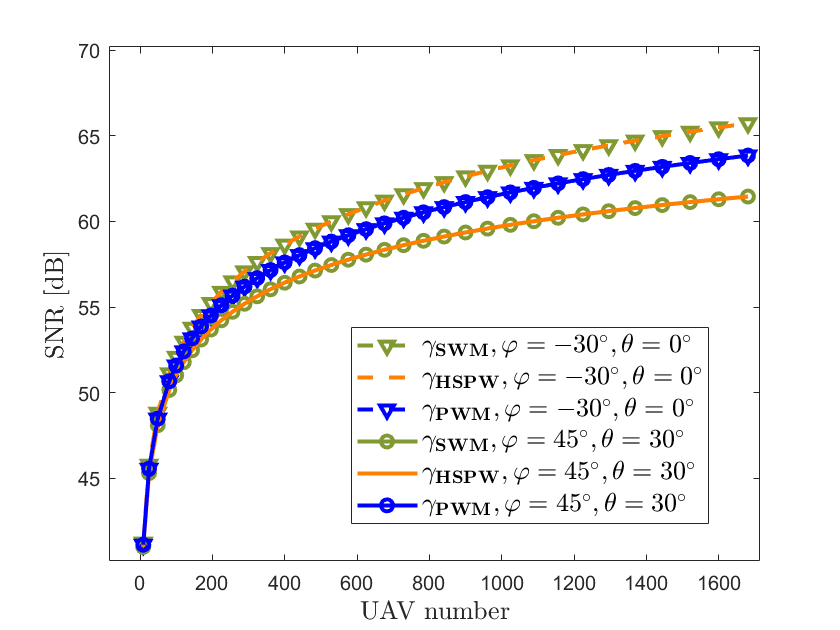}
\caption{SNRs versus the number of array elements with different models and different $\varphi$ and $\theta$.}
\label{fig_3}
\vspace{-0.25cm}
\end{figure}

\section{Channel Estimation}
In this section, leveraging the inherent characteristics of the distributed near-field scenario of multi-UAV, we design effective near-field channel estimation methods utilizing the accurate SWM and its approximation HSPWM, i.e., SD-OMP and tensor-OMP. In our approach, we assume that each UAV receives the user’s pilot signal using its onboard analog combiner, and the resulting low-dimensional compressed signals are subsequently shared with a central fusion node (e.g., a leader UAV or ground station) for joint processing. This centralized assumption enables high-accuracy channel estimation and simplifies algorithm design. Investigation of distributed estimation schemes and the effects of practical inter-UAV communication constraints will be considered in future work.

\subsection{SD-OMP based on SWM}
In this section, we propose a sparse estimation method based on the SWM. This method extends the angle-distance estimation framework from the PD to the SD, enabling the estimation of elevation, azimuth, and distance, and thus generating the SD dictionary.
Moreover, compressive sensing techniques are employed to extract the essential angle and distance parameters necessary for channel reconstruction.

Specifically, the dictionary representation of channel can be formulated as 
\begin{small}
\begin{equation}\label{rec}
\hat{\boldsymbol{h}} \approx \boldsymbol{Gz},
\vspace{-3mm}
\end{equation}
\end{small}
where
\begin{small}
\begin{equation}
\begin{aligned}
&\boldsymbol{G} = \left[ \boldsymbol{g}_{\mathbf{SWM}}\left( \theta_1, \varphi_1,r_1\right), \ldots, \boldsymbol{g}_{\mathbf{SWM}}\left(\theta_A, \varphi_1, r_1 \right), \right.\\ &\left.
\boldsymbol{g}_{\mathbf{SWM}}\left(\theta_A, \varphi_B, r_1 \right), \ldots, \boldsymbol{g}_{\mathbf{SWM}}\left(\theta_A, \varphi_B,r_C\right) \right]\in \mathbb{C}^{MN \times U }, 
\end{aligned}
\end{equation}
\end{small}
$\left\{\theta_1,\ldots,\theta_A\right\}$, $\left\{\varphi_1,\ldots,\varphi_B\right\}$ and $\left\{r_1,\ldots, r_C\right\}$ denote the sampling points for the azimuth angle $\theta$, the elevation angle $\varphi$ and the distance $r$, respectively. $\boldsymbol{G}$ denotes the SD dictionary, and $U=ABC$ represents the number of atoms.
$\boldsymbol{z}\in \mathbb{C}^{U \times 1 }$ is a sparse vector, which physically signifies the reconstruction of $\boldsymbol{h}$ using the minimal number of non-zero values. The non-zero values represent the channel path gains, and these non-zero value indices indicate the selected atoms for the channel reconstruction.

The problem is transformed into minimizing the number of non-zero entries of $\boldsymbol{z}$, hence we have
\begin{small}
\begin{equation}\label{1D}
\setlength{\abovedisplayskip}{3pt}
\setlength{\belowdisplayskip}{3pt}
\begin{aligned}
\underset{\boldsymbol{z}}{\arg\min} & \quad \|\boldsymbol{z}\|_0 \\[-3mm]
\text{s.t.} & \quad \left\|\boldsymbol{y} - \boldsymbol{W}^H \boldsymbol{G} \boldsymbol{z}\right\|_2^2 \leq \epsilon
\end{aligned}
\end{equation}
\end{small}
where $\epsilon$ denotes the precise factor.

The problem proposed in (\ref{1D}) can be addressed by integrating it with the traditional OMP algorithm for sparse signal recovery, namely SD-OMP-ongrid. Due to its universality in many literatures like \cite{NF1}, a detailed explanation has been omitted. 

After recovering $\boldsymbol{z}$, the restored channel $\hat{\boldsymbol{h}}$ can be obtained according to (\ref{rec}). The channel parameters of all paths can be obtained by $(\widehat{\boldsymbol{\theta}},\widehat{\boldsymbol{\varphi}},\widehat{\boldsymbol{r}}) = [(\widehat{\theta}_1, \widehat{\varphi}_1, \widehat{r}_1), \ldots, (\widehat{\theta}_L, \widehat{\varphi}_L,\widehat{r}_L)]$.

Furthermore, to achieve more accurate channel parameters,  a new optimization problem can be formulated as
\begin{small}
\begin{equation}\label{LSS}
\setlength{\abovedisplayskip}{3pt}
\setlength{\belowdisplayskip}{3pt}
(\widetilde{\widehat{\boldsymbol{\theta}}},\widetilde{\widehat{\boldsymbol{\varphi}}},\widetilde{\widehat{\boldsymbol{r}}}) =\arg\min_{(\widehat{\boldsymbol{\theta}},\widehat{\boldsymbol{\varphi}},\widehat{\boldsymbol{r}})}\left\|  \boldsymbol{y}-(\boldsymbol{W}^{\text{H}} \widehat{\boldsymbol{G}}) (\boldsymbol{W}^{\text{H}} \widehat{\boldsymbol{G}})^{\dagger} \boldsymbol{y}\right\|.
\end{equation}
\end{small}
where 
\begin{small}
\begin{equation}
\begin{aligned}
\widehat{\boldsymbol{G}}=&\left[ \boldsymbol{g}_{\mathbf{SWM}}\left( \theta_1, \varphi_1,r_1\right), \ldots, \boldsymbol{g}_{\mathbf{SWM}}\left(\theta_l, \varphi_l, r_l \right), \right. \\ &\left. \ldots, \boldsymbol{g}_{\mathbf{SWM}}\left(\theta_L, \varphi_L,r_L\right) \right]\in \mathbb{C}^{MN \times L }
\end{aligned}
\end{equation}
\end{small}
is the estimation result by SD-OMP-ongrid.

The Nelder-Mead algorithm \cite{NMdanchuan}, a derivative-free method, is utilized to solve (\ref{LSS}) by taking the SD-OMP-ongrid results as initial values. This approach, termed SD-OMP-offgrid, enables lightweight local refinement in low-dimensional angular space without requiring gradient information. It is particularly suitable for our non-linear near-field model with array perturbations, where analytical gradients are difficult to obtain.

\subsection{Tensor-OMP based on HSPWM}
The methods based on SD-OMP provide an effective solution for multi-UAV near-field channel estimation. However, due to their high computational complexity required by the large-scale dictionary, these methods do not fully leverage the characteristics of multi-UAV near-field scenario, thereby not suitable for distributed ELAA systems. As a result, we propose a sparse estimation method based on the HSPWM in this subsection. Leveraging the inherent properties of the HSPWM, the near-field
array response can be represented as
the outer product of three vectors. It expands the PD dictionary to encompass the tensor dictionary, capitalizing on its distinctive mathematical formulation to simplify the representation of the dictionary.
\subsubsection{Tensor dictionary} Inspired by \cite{3DOMP},
the construction of the tensor atoms is necessary. Reviewing (\ref{afj}), it can be readily derived that
\begin{small}
\begin{equation}\label{Afj}
\setlength{\abovedisplayskip}{-2pt}
\begin{aligned}
\boldsymbol{a}(\theta, \varphi) &= \sqrt{\beta_0}  \boldsymbol{u}(\theta, \varphi) \otimes \boldsymbol{v}(\theta, \varphi) ,
\end{aligned}
\end{equation}
\end{small}
where $\boldsymbol{u}(\theta, \varphi) = \left[1, \ldots, e^{j \frac{2\pi}{\lambda} d(N_x-1) sin\varphi sin\theta }\right]\in \mathbb{C}^{N_x \times 1 }$, $\boldsymbol{v}(\theta, \varphi) = \left[1, \ldots, e^{j \frac{2\pi}{\lambda} d(N_y-1) sin\varphi cos\theta }\right]\in \mathbb{C}^{N_y \times 1 }$. 

Combining (\ref{C}) and (\ref{Afj}), it is evident that $\boldsymbol{g}_{\mathbf{HSPWM}}(\theta, \varphi, r)$ can be represented by the Kronecker product of three vectors, which can be reshaped into a tensor with a closed-form expression. According to the rule of tensor vectorization, it is given by 
\begin{small}
\begin{equation} 
\begin{aligned}
\text{devec}(\boldsymbol{g}_{\mathbf{HSPWM}}(\theta, \varphi, r)) 
&=\boldsymbol{\mathcal{G}}(\theta, \varphi, r)\\[-1mm]&
=\text{devec}(\boldsymbol{b}(\theta, \varphi, r)  \otimes \boldsymbol{u}(\theta, \varphi)  \otimes \boldsymbol{v}(\theta, \varphi))
\\[-1mm]&= \boldsymbol{v}(\theta, \varphi)  \circ \boldsymbol{u}(\theta, \varphi)    \circ \boldsymbol{b}(\theta, \varphi, r),
\end{aligned}
\vspace{-0.1cm}
\end{equation}
\end{small}
where "$\circ$" denote the outer product, $\boldsymbol{\mathcal{G}}(\theta, \varphi, r)$ is a third-order tensor with the size of $N_y \times N_x \times M$ and devec(·) denotes the devectorization operator.

Consequently, the tensor-form channel can be expressed as:
\begin{small}
\begin{equation}
\setlength{\abovedisplayskip}{1pt}
\setlength{\belowdisplayskip}{0.1pt}
\begin{aligned}
\text{devec}(\boldsymbol{h}) &= \boldsymbol{\mathcal{H}}\\[-2mm]
&\approx \frac{1}{\sqrt{L} }\sum_{l=1}^{L} z_{l} \boldsymbol{\mathcal{G}}(\theta_{l},\varphi_{l}, r_{l}) \\[-2mm]
&= \frac{1}{\sqrt{L} } \sum_{l=1}^{L} z_{l} \boldsymbol{v}(\theta_l, \varphi_l) \circ \boldsymbol{u}(\theta_l, \varphi_l)    \circ \boldsymbol{b}(\theta_l, \varphi_l, r_l).
\end{aligned}
\end{equation}
\end{small}

The dictionary representation of tensor-form channel reconstruction can be formulated as:
\begin{small}
\begin{equation}\label{Hre}
\begin{aligned}
\hat{\boldsymbol{\mathcal{H}}} = \sum_{a \in \{1, \ldots, A\}} 
\sum_{b \in \{1, \ldots, B\}} \sum_{c \in \{1, \ldots, C\}} \kappa_{a,b,c} \boldsymbol{\mathcal{T}}_{a,b,c},
\end{aligned}
\end{equation}
\end{small}
where $\boldsymbol{\mathcal{T}}\in \mathbb{C}^{A\times B \times C \times N_y \times N_x \times M } $ is a sixth-order tensor, which constitutes the tensor dictionary. $\boldsymbol{\mathcal{T}}_{a,b,c}=\boldsymbol{\mathcal{G}}(\theta_{a},\varphi_{b}, r_{c})=\boldsymbol{v}(\theta_a, \varphi_b) \circ \boldsymbol{u}(\theta_a, \varphi_b)    \circ \boldsymbol{b}(\theta_a, \varphi_b, r_c)
 \in \mathbb{C}^{N_y \times N_x \times M }$ is the $(a,b,c)$-th atom, where $\theta_a$, $\varphi_b$ and $r_c$ denote the $a$-th azimuth angle sample, the $b$-th elevation angle sample and the $c$-th distance sample of the tensor dictionary. $\boldsymbol{\kappa}$ is used to store $\kappa_{a,b,c}$, which is a sparse vector. If the value of $\kappa_{a,b,c}$ is non-zero, the $(a,b,c)$-th atom from the tensor dictionary is selected and the corresponding value represents the channel gain.

According to (\ref{RS}), the received signal can be rewritten as
\begin{small}
\begin{equation}
\setlength{\abovedisplayskip}{1.5pt}
\setlength{\belowdisplayskip}{1.5pt} 
\begin{aligned}
    \boldsymbol{y} =& \left(\mathbf{I}_{M} \otimes \boldsymbol{W}_x^H \otimes \boldsymbol{W}_y^H\right) \text{vec}(\boldsymbol{\mathcal{H}})\\[-2mm]
    =&\frac{1}{\sqrt{L} } \sum_{l=1}^{L} z_{l}\left[\left(\mathbf{I}_{M} \otimes \boldsymbol{W}_x^H \otimes \boldsymbol{W}_y^H\right) \right.\\[-1mm] &\left.
    \left(\boldsymbol{b}(\theta_l, \varphi_l, r_l)\otimes \boldsymbol{u}(\theta_l, \varphi_l)    \otimes 
\boldsymbol{v}(\theta_l, \varphi_l)  \right) \right] \\[-1mm]
=&\frac{1}{\sqrt{L} } \sum_{l=1}^{L} z_{l}\left( \boldsymbol{b}(\theta_l, \varphi_l, r_l)\otimes \overline{\boldsymbol{u}}(\theta_l, \varphi_l) \otimes \overline{\boldsymbol{v}}(\theta_l, \varphi_l)\right),
\end{aligned}
\vspace{-1mm}
\end{equation}
\end{small}
where $\overline{\boldsymbol{u}}(\theta_l, \varphi_l)=\boldsymbol{W}_x^H \boldsymbol{u}(\theta_l, \varphi_l)\in \mathbb{C}^{Q_x \times 1}$, $\overline{\boldsymbol{v}}(\theta_l, \varphi_l)=\boldsymbol{W}_y^H  
\boldsymbol{v}(\theta_l, \varphi_l)\in \mathbb{C}^{Q_y \times 1}$. The result is obtained using $(\mathbf{A} \otimes \mathbf{B})(\mathbf{C} \otimes \mathbf{D}) = \mathbf{AC} \otimes \mathbf{BD}$. Additionally, vec(·) denotes the vectorization operator.

Employing the devectorization operation, the tensor-form received signal can be expressed as
\begin{small}
\begin{equation}\label{MY}
\begin{aligned}
\boldsymbol{\mathcal{Y}}
&=\text{devec}(\boldsymbol{y})\\[-1mm]&
=\frac{1}{\sqrt{L} } \sum_{l=1}^{L} z_{l}\overline{\boldsymbol{v}}(\theta_l, \varphi_l) \circ \overline{\boldsymbol{u}}(\theta_l, \varphi_l) \circ \boldsymbol{b}(\theta_l, \varphi_l, r_l),   
\end{aligned}
\end{equation}
\end{small}
where $\boldsymbol{\mathcal{Y}}\in \mathbb{C}^{Q_y\times Q_x \times M }$ is the third-order tensor.

According to (\ref{Hre}) and (\ref{MY}), we define $\overline{\boldsymbol{\mathcal{WT}}}_{a,b,c}=\overline{\boldsymbol{v}}(\theta_a, \varphi_b) \circ \overline{\boldsymbol{u}}(\theta_a, \varphi_b) \circ \boldsymbol{b}(\theta_a, \varphi_b, r_c)  $. Therefore, the dictionary representation of the received signal can be expressed as 
\begin{small}
\begin{equation}\label{Y^}
\begin{aligned}
\hat{\boldsymbol{\mathcal{Y}}} &=\sum_{a \in \{1, \ldots, A\}} 
\sum_{b \in \{1, \ldots, B\}} \sum_{c \in \{1, \ldots, C\}}
\kappa_{a,b,c} 
\overline{\boldsymbol{\mathcal{WT}}}_{a,b,c}.
\end{aligned}
\vspace{-1mm}
\end{equation}
\end{small}

The channel reconstruction problem is then transformed into a tensor recovery issue utilizing a tensor dictionary, which is given by
 \begin{small}
\begin{equation}\label{3D}
\begin{aligned}
    \arg\min_{\boldsymbol{\kappa}} & \quad \|\boldsymbol{\kappa}\|_0 \\[-2mm]
\text{s.t.} & \quad \left\| \boldsymbol{\mathcal{Y}} - \sum_{a \in \{1, \ldots, A\}} \sum_{b \in \{1, \ldots, B\}} 
\sum_{c \in \{1, \ldots, C\}}\right.\\ &\left. \kappa_{a,b,c} \overline{\boldsymbol{\mathcal{WT}}}_{a,b,c} \right\|_F^2 \leq \epsilon. 
\end{aligned}
\vspace{-1.5mm}
\end{equation}
 \end{small}

After recovering $\boldsymbol{\kappa}$, the restored channel $\hat{\boldsymbol{\mathcal{H}}}$ can be obtained
according to (\ref{Hre}).

Since (\ref{3D}) requires computation with a tensor dictionary, its recovery process deviates from the standard method outlined in (\ref{1D}). In this subsection, we present a greedy solution that includes the selection of the tensor atom to tackle this distinctive challenge.

\subsubsection{3D Projection}
The projection of the received signal $\boldsymbol{\mathcal{Y}}$ onto the third-order tensor is defined as:
 \begin{small}
\begin{equation}
  \frac{ \left\langle  \boldsymbol{\mathcal{Y}}, \overline{\boldsymbol{\mathcal{WT}}}_{a,b,c} \right\rangle }{\left\| \overline{\boldsymbol{\mathcal{WT}}}_{a,b,c} \right\|_2},
\end{equation}
 \end{small}
where 
 \begin{small}
\begin{equation}
\begin{aligned}
\left\langle  \boldsymbol{\mathcal{Y}}, \overline{\boldsymbol{\mathcal{WT}}}_{a,b,c} \right\rangle=\boldsymbol{\mathcal{Y}}\times_1 
\overline{\boldsymbol{v}}(\theta_a, \varphi_b)
  \times_2\\\overline{\boldsymbol{u}}(\theta_a, \varphi_b) \times_3  \boldsymbol{b}(\theta_a, \varphi_b, r_c),
\end{aligned}
\vspace{-2mm}
\end{equation}
 \end{small}
  \begin{small}
\begin{equation}
\begin{aligned}
&\left\| \overline{\boldsymbol{\mathcal{WT}}}_{a,b,c} \right\|_2\\
&=\sqrt{\sum_{m=1}^{M}\sum_{q_x=1}^{Q_x}\sum_{q_y=1}^{Q_y} \left(  \overline{v}_{q_y} (\theta_a, \varphi_b) \overline{u}_{q_x}(\theta_a, \varphi_b)  b_{m}(\theta_a, \varphi_b, r_c)  \right)^2} \\
&= \left\| \overline{\boldsymbol{v}}(\theta_a, \varphi_b) \right\|_2   \left\| \overline{\boldsymbol{u}}(\theta_a, \varphi_b) \right\|_2 \left\| \boldsymbol{b}(\theta_a, \varphi_b, r_c) \right\|_2
\end{aligned}
\vspace{-0.2cm}
\end{equation}
 \end{small}
with $\times_1, \times_2, \times_3$ denoting the $n$-mode product of a tensor with the vector.  

\begin{figure}[!t]
\setlength{\abovecaptionskip}{-0.1cm} 
\centering
\includegraphics[width=3in] {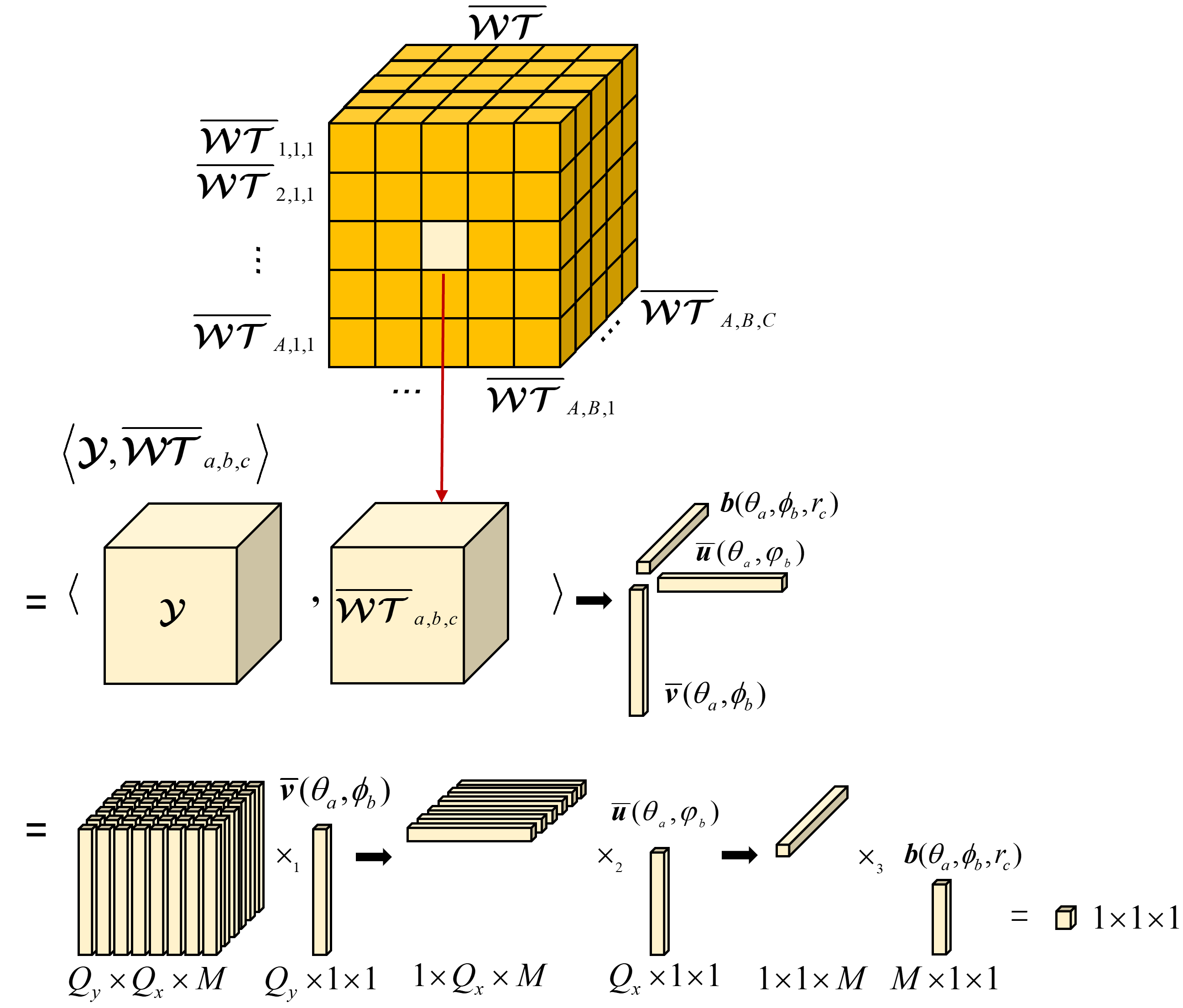}
\caption{An intuitive visualization of the tensor multiplication process.}
\label{fig_4}
\vspace{-0.25cm} 
\end{figure}

A deeper understanding of the tensor multiplication process can be achieved by Fig. 4, which a grid represents a third-order tensor. The candidate atoms of Fig. 4 are extracted and reassembled into $\overline{\boldsymbol{\mathcal{WT}}}$, with dimensions of length 
$A$, width $B$, and height $C$, thus forming a sixth-order tensor. A particular element within $\overline{\boldsymbol{\mathcal{WT}}}$, denoted as 
$\overline{\boldsymbol{\mathcal{WT}}}_{a,b,c}$
 , has dimensions of
$Q_y \times Q_x \times M $. The multiplication of 
$\boldsymbol{\mathcal{Y}}$
 with 
$\overline{\boldsymbol{\mathcal{WT}}}_{a,b,c}$ can be regarded as the sequential mode products of the tensor with three vectors. Each mode product reduces the order of the tensor by one, ultimately transforming it into a zeroth-order tensor, which is a scalar. The final result consists of $ABC$ values. Identifying the indices of the maximum value among these values corresponds to the angles and distances that we are seeking to determine.

\begin{algorithm}[!t]
\small
\caption{Tensor-OMP-ongrid/offgrid Procedure}\label{alg:alg1}
\begin{algorithmic}
\STATE
\STATE$ \textbf{Inputs:}$ 
\begin{itemize}
    \item $\boldsymbol{W}_x \in \mathbb{C}^{N_x \times Q_x}$: sampling matrix
    \item $\boldsymbol{W}_y \in \mathbb{C}^{N_y \times Q_y}$: sampling matrix
    \item $\boldsymbol{\mathcal{Y}} \in \mathbb{C}^{Q_y\times Q_x \times M }$: tensor-form received signal 
  \item  $\boldsymbol{\mathcal{T}}\in \mathbb{C}^{A\times B \times C \times N_y \times N_x \times M }$: dictionary
    \item $\boldsymbol{\mathcal{R}}^{(l)}$: residual
    \item $L$: sparsity level
    \item $\Gamma$:the set of selected atom indices
\end{itemize}
\STATE$ \textbf{Outputs:}$ 
\begin{itemize}
    \item $\widehat{\boldsymbol{h}}$: the reconstructed channel by tensor-OMP-ongrid
    \item 
$\widetilde{\widehat{\boldsymbol{h}}}$: the reconstructed channel by tensor-OMP-offgrid
\end{itemize}
\STATE $ \textbf{Begin}$
\begin{itemize}
    \item $\boldsymbol{\mathcal{R}}^{(0)} = \boldsymbol{\mathcal{Y}}$
    \item $(a, b, c) \leftarrow \{(1, 1, 1), (1, 1, 2), \ldots, (A, B, C)\}$
    \item $\Gamma = \emptyset$
\end{itemize}
\FOR{$l = 1, \ldots, L$}
    \STATE $(a^*,b^*,c^*) = \arg\max_{a,b,c} \frac{ \left\langle  \boldsymbol{\mathcal{R}}, \overline{\boldsymbol{\mathcal{WT}}}_{a,b,c} \right\rangle }{\left\| \overline{\boldsymbol{\mathcal{WT}}}_{a,b,c} \right\|_2}$;
    \STATE $\Gamma \leftarrow \Gamma \cup (a^*,b^*, c^*)$,$(a, b, c) \leftarrow (a, b, c) \setminus (a^*,b^*, c^*);$
    \STATE $\boldsymbol{X} \leftarrow [\boldsymbol{X}, vec(\overline{\boldsymbol{\mathcal{WT}}}_{a^*,b^*,c^*})]$;
    \STATE $\boldsymbol{\kappa}_{\Gamma} = \left( \boldsymbol{X}^H \boldsymbol{X} \right)^{-1} \boldsymbol{X}^H vec\left( \boldsymbol{\mathcal{R}}^{(l)} \right)$;
    \STATE $\boldsymbol{\mathcal{R}}^{(l)} = \boldsymbol{\mathcal{R}}^{(l)} - \sum_{(a,b, c) \in \Gamma} \kappa_{a,b, c} \overline{\boldsymbol{\mathcal{WT}}}_{a,b,c}$;
\ENDFOR
\STATE \%\% Ongrid
\STATE $\boldsymbol{\kappa}_{\Gamma} = \left( \boldsymbol{X}^H \boldsymbol{X} \right)^{-1} \boldsymbol{X}^H \text{vec}\left( \boldsymbol{\mathcal{Y}} \right)$;
\STATE 
$\widehat{\boldsymbol{\mathcal{H}}} =\sum_{(a,b, c) \in \Gamma} \kappa_{a,b, c} \boldsymbol{\mathcal{T}}_{a,b,c}$
\STATE
$\widehat{\boldsymbol{h}} = \text{devec}(\widehat{\boldsymbol{\mathcal{H}}})$;
\STATE \%\% Offgrid
\STATE Estimate $(\widetilde{\widehat{\boldsymbol{\theta}}},\widetilde{\widehat{\boldsymbol{\varphi}}},\widetilde{\widehat{\boldsymbol{r}}})$ by (\ref{uLL});
\STATE
$\widetilde{\widehat{\boldsymbol{h}}} = \widetilde{\widehat{\boldsymbol{G}}} \left( \boldsymbol{{W}}^{\text{H}} \widetilde{\widehat{\boldsymbol{G}} }\right)^{\dagger} \text{vec}\left( \boldsymbol{\mathcal{Y}}\right)$;
\RETURN $\widetilde{\widehat{\boldsymbol{h}}}$
\end{algorithmic}
\label{alg1}
\end{algorithm}

\subsubsection{Least squares}
The set of selected atom indices is defined as $\Gamma$. These selected atoms can be used to construct an approximation of 
$\boldsymbol{\mathcal{Y}}$, that is:
 \begin{small}
\begin{equation}
\setlength{\abovedisplayskip}{3pt}
\setlength{\belowdisplayskip}{3pt}
\begin{aligned}
\boldsymbol{\mathcal{Y}}_{\Gamma} = \sum_{(a, b,c) \in \Gamma} \kappa_{a, b,c} \overline{\boldsymbol{\mathcal{WT}}}_{a,b,c}. 
\end{aligned}
\vspace{-0.2cm}
\end{equation}
 \end{small}

We define a tensor $\boldsymbol{\mathcal{R}}\in \mathbb{C}^{Q_y \times Q_x \times M}$ to represent the residual: 
$\boldsymbol{\mathcal{R}}=\boldsymbol{\mathcal{Y}}-\boldsymbol{\mathcal{Y}}_{\Gamma}$. By seeking the optimal 
$\boldsymbol{\kappa}_{\Gamma}$ to minimize the Frobenius norm of $\boldsymbol{\mathcal{R}}$, the problem is essentially equivalent to a least squares problem. Consequently, the optimization problem can be formulated as
 \begin{small}
\begin{equation}
\setlength{\abovedisplayskip}{-0.1pt}
\setlength{\belowdisplayskip}{-0.1pt}
\begin{aligned}
\arg\min_{\boldsymbol{\kappa}_{\Gamma}} \left\| \boldsymbol{\mathcal{Y}} - \sum_{(a, b,c) \in \Gamma} \kappa_{a, b,c} \overline{\boldsymbol{\mathcal{WT}}}_{a,b,c} \right\|_F^2.
\end{aligned}
\end{equation}
 \end{small}

Unlike the traditional residual computation method used in OMP , we need to solve for the tensor-form coefficients, which undoubtedly poses a challenge to the solution process. This problem can be simplified through the operation of vectorization. Firstly, we have
 \begin{small}
\begin{equation}
\setlength{\abovedisplayskip}{-0.1pt}
\setlength{\belowdisplayskip}{-0.1pt}
\begin{aligned}
&\text{vec}\left(\sum_{(a, b,c)\in\Gamma}\kappa_{a,b, c} \overline{\boldsymbol{\mathcal{WT}}}_{a,b,c}\right) \\ &= \sum_{(a,b, c)\in\Gamma}\kappa_{a,b, c} \text{vec}\left(\overline{\boldsymbol{\mathcal{WT}}}_{a,b,c}\right) \\
&= \boldsymbol{X} \boldsymbol{\kappa}_{\Gamma},
\end{aligned}
\end{equation}
\end{small}
where $\boldsymbol{X} \in \mathbb{C}^{MQ \times length(\Gamma)}$ is constructed by the Kronecker product of three vectors,  and $length(\Gamma)$ denotes the number of selected atom.

Then the tensor-form minimization problem can be simplified to
 \begin{small}
\begin{equation}
\setlength{\abovedisplayskip}{-0.1pt}
\setlength{\belowdisplayskip}{-0.1pt}
\begin{aligned}
\arg\min_{\boldsymbol{\kappa}_{\Gamma}} \left\| \text{vec}(\boldsymbol{\mathcal{Y}}) - \boldsymbol{X} \boldsymbol{\kappa}_{\Gamma} \right\|_2^2.
\end{aligned}
\vspace{-1.5mm}
\end{equation}
 \end{small}

$\boldsymbol{\kappa}_{\Gamma}$ can be calculated as
 \begin{small}
\begin{equation}
\setlength{\abovedisplayskip}{-0.1pt}
\setlength{\belowdisplayskip}{-0.1pt}
\begin{aligned}
\boldsymbol{\kappa}_{\Gamma} = \left( \boldsymbol{X}^H \boldsymbol{X} \right)^{-1} \boldsymbol{X}^H \text{vec}\left( \boldsymbol{\mathcal{Y}} \right).
\end{aligned}
\vspace{-1mm}
\end{equation}
 \end{small}

After $\boldsymbol{\kappa}_{\Gamma}$ is obtained, the residual $\boldsymbol{\mathcal{R}}$ can be obtained as 
\begin{small}
\begin{equation}
\begin{aligned}
\boldsymbol{\mathcal{R}}^{(l)} = \boldsymbol{\mathcal{Y}} - \sum_{(a,b, c) \in \Gamma} \kappa_{a,b, c} \overline{\boldsymbol{\mathcal{WT}}}_{a,b,c}.
\end{aligned}
\vspace{-0.1cm}
\end{equation}
\end{small}

\subsubsection{Offgrid}
The aforementioned procedures have been summarized into Algorithm 1, which is called tensor-OMP-ongrid. 

Building upon these foundations, the pursuit of enhanced channel estimation precision leads us to further optimize the method. This advancement is realized through the tensor-OMP-offgrid, which evolves from the outcomes of the tensor-OMP-ongrid algorithm. Specifically, a new optimization problem can be formulated as
 \begin{small}
\begin{equation}\label{uLL}
\setlength{\abovedisplayskip}{1pt}
\setlength{\belowdisplayskip}{1pt}
(\widetilde{\widehat{\boldsymbol{\theta}}},\widetilde{\widehat{\boldsymbol{\varphi}}},\widetilde{\widehat{\boldsymbol{r}}}) =\arg\min_{(\widehat{\boldsymbol{\theta}},\widehat{\boldsymbol{\varphi}},\widehat{\boldsymbol{r}}))}\left\| \text{vec}\left( \boldsymbol{\mathcal{Y}}\right)-(\boldsymbol{W}^{\text{H}} \widehat{\boldsymbol{G}}) (\boldsymbol{W}^{\text{H}} \widehat{\boldsymbol{G}})^{\dagger} \text{vec}\left( \boldsymbol{\mathcal{Y}}\right)\right\|,
\vspace{-2mm}
\end{equation}
 \end{small}
where 
 \begin{small}
\begin{equation}
\begin{aligned}
\widehat{\boldsymbol{G}}=&\left[ \boldsymbol{g}_{\mathbf{SWM}}\left( \theta_1, \varphi_1,r_1\right), \ldots, \boldsymbol{g}_{\mathbf{SWM}}\left(\theta_l, \varphi_l, r_l \right), \right. \\ &\left. \ldots, \boldsymbol{g}_{\mathbf{SWM}}\left(\theta_L, \varphi_L,r_L\right) \right]\in \mathbb{C}^{MN \times L }
\end{aligned}
\end{equation}
 \end{small}
is the estimation result by the tensor-OMP-ongrid.

The Nelder-Mead algorithm is utilized to solve equation (\ref{uLL}) by taking the results of tensor-OMP-ongrid as initial
values. The detailed procedural steps for this refined approach are delineated comprehensively within Algorithm 1.

\subsection{Disscussion}
\subsubsection{Sparsity discussion}
(\ref{3D}) is undoubtedly a sparse recovery problem, which can be understood through Fig. 5.

In Fig. 5, each grid represents a third-order tensor, which corresponds to a atom. We first construct $\overline{\boldsymbol{U}} = \left[ \overline{\boldsymbol{u}}(\theta_1, \varphi_1), \ldots, \overline{\boldsymbol{u}}(\theta_A, \varphi_B) \right] \in \mathbb{C}^{N_x \times AB}, 
\overline{\boldsymbol{V}} = \left[ \overline{\boldsymbol{v}}(\theta_1, \varphi_1), \ldots, \overline{\boldsymbol{v}}(\theta_A, \varphi_B) \right] \in \mathbb{C}^{N_y \times AB},
\boldsymbol{B} = \left[ \boldsymbol{b}(\theta_1, \varphi_1, r_1), \ldots, \boldsymbol{b}(\theta_A, \varphi_B, r_C) \right] \in \mathbb{C}^{M \times ABC}$,
from which it can be inferred that the size of the cubic structure is 
$AB\times AB\times ABC$. This also implies that the sampling required would be $A^3B^3C$, which is substantially excessive. However, it is evident that due to the consistency of $\theta$ and $\varphi$ across the three vectors $\overline{\boldsymbol{u}}$, $\overline{\boldsymbol{v}}$ and $\boldsymbol{b}$, the candidate atoms will only appear in specific positions, denoted by 
$s=x=a$, $t=y=b$. These positions are represented by the yellow grids in Fig. 5, which in this context signify  $\overline{\boldsymbol{\mathcal{WT}}}_{a,b,c}$. Notably, these yellow grids lie along the \emph{body diagonal} of the tensor cube, ensuring the angular consistency $(\theta_a, \varphi_b)$ across the three components $\overline{\boldsymbol{u}}$, $\overline{\boldsymbol{v}}$, and $\boldsymbol{b}$. 
The ellipsis ``$\cdots$'' along the third dimension denotes the continuation over the range index $r_c$, showing that only atoms with angular alignment contribute to the sparse representation. 
The remaining blue grids indicate invalid or impossible atoms that violate the angular consistency rule, e.g., $
\overline{\boldsymbol{v}}(\theta_1, \varphi_4) \circ \overline{\boldsymbol{u}}(\theta_8, \varphi_2) \circ \boldsymbol{b}(\theta_3, \varphi_7, r_1),$
which combine unmatched angles and hence are excluded from the valid dictionary. 
This design effectively avoids redundant angular sampling and highlights the inherent sparsity of the problem.

\begin{figure}[!t]
\setlength{\abovecaptionskip}{-0.2cm} 
\centering
\includegraphics[width=3.5in] {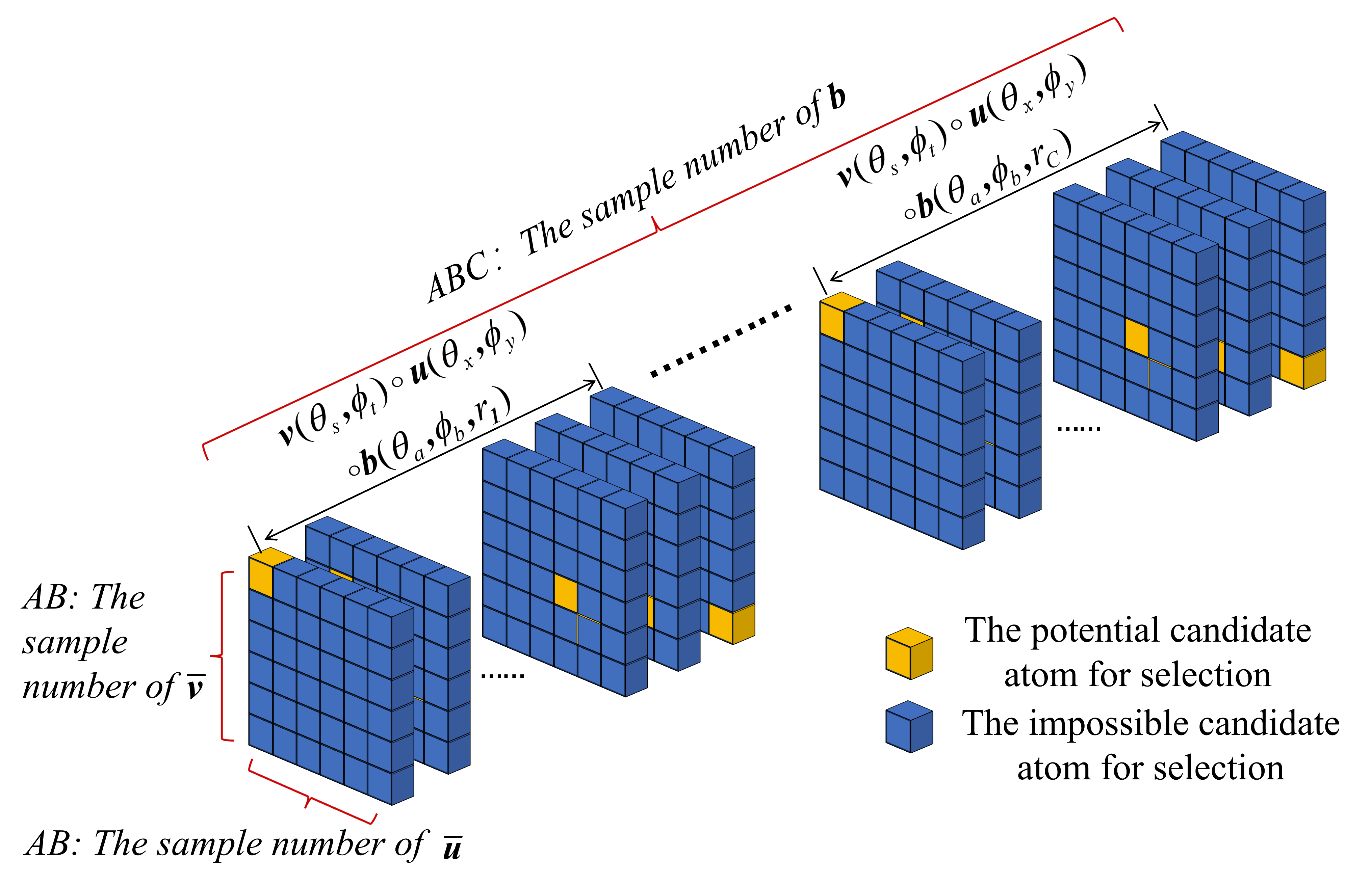}
\caption{An intuitive visualization of the unique sparsity.}
\label{fig_5}
\vspace{-0.2cm}
\end{figure}

\subsubsection{Complexcity analysis} In SD-OMP, the projection step requires multiplying the large matrix 
$\boldsymbol{W}^H \boldsymbol{G}
\in \mathbb{C}^{MQ \times U}$
  and computing the inner product with the residual vector.
 The projection step requires computing the matrix product $\boldsymbol{W}^H \boldsymbol{G} \in \mathbb{C}^{MQ \times ABC}$, resulting in a complexity of $\mathcal{O}(NM\cdot MQ \cdot U)$. In contrast, Tensor-OMP avoids explicit dictionary construction and performs three mode-$n$ tensor products per atom, with each projection costing only $\mathcal{O}(Q_x Q_y M)$. Therefore, the total projection complexity is reduced to $\mathcal{O}( M Q\cdot U)$. This approach leverages the Kronecker structure and compressive projections, significantly reducing computational cost. While existing OMP variants such as SDAR-OMP \cite{SDAR} and 2D-OMP \cite{2DOMP} have been proposed to reduce the complexity of sparse recovery, they differ significantly from our tensor-OMP method.
Specifically, successive decision-aided recovery (SDAR)-OMP recovers multi-dimensional sparse vectors in a sequential manner under the Kronecker sensing model, which requires intermediate decision steps and lacks global tensor structure exploitation. 2D-OMP, on the other hand, operates in the matrix domain and is limited to 2D signal recovery.
In contrast, our method leverages the structure of the near-field HSPWM array to define tensor atoms, and performs matching directly in the tensor domain, enabling batch projection onto the entire dictionary with significantly lower complexity.

\begin{figure*}[!t]
 \vspace{-1.2cm} 
\centering
\subfloat[]{\includegraphics[width=0.39\linewidth]{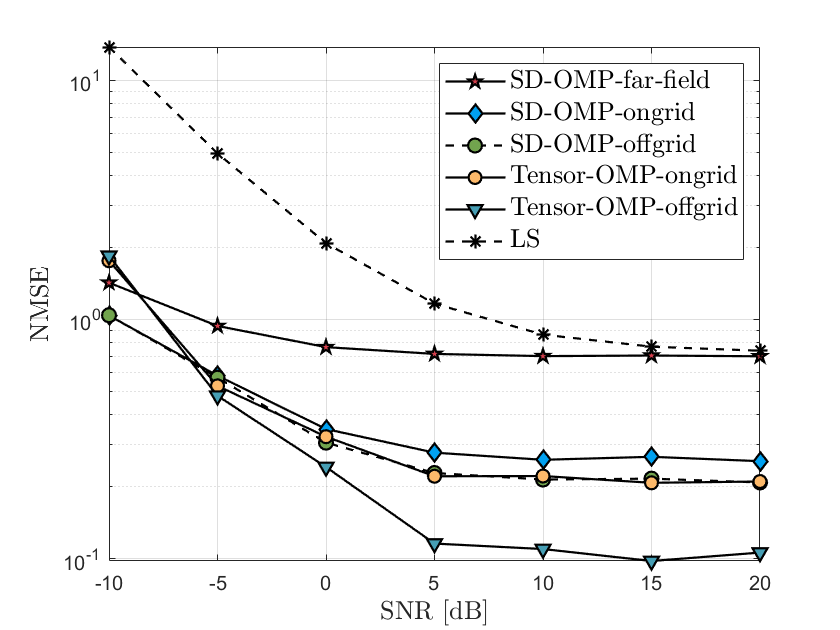}%
\label{fig6_first_case}}
\hfil
\subfloat[]{\includegraphics[width=0.39\linewidth]{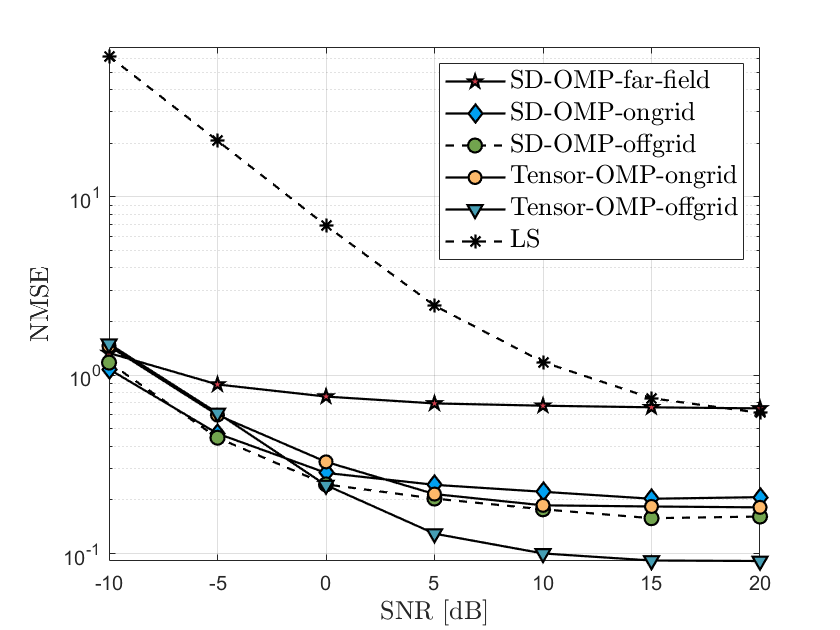}%
\label{fig6_second_case}}
\hfil
\\
\vspace{-0.18in}
\subfloat[]{\includegraphics[width=0.39\linewidth]{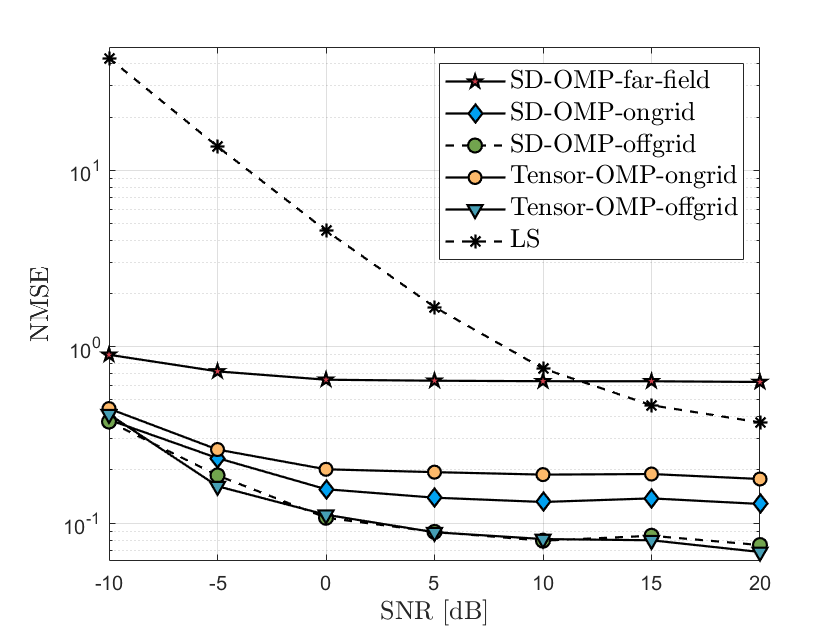}%
\label{fig6_third_case}}
\hfil
\subfloat[]{\includegraphics[width=0.39\linewidth]{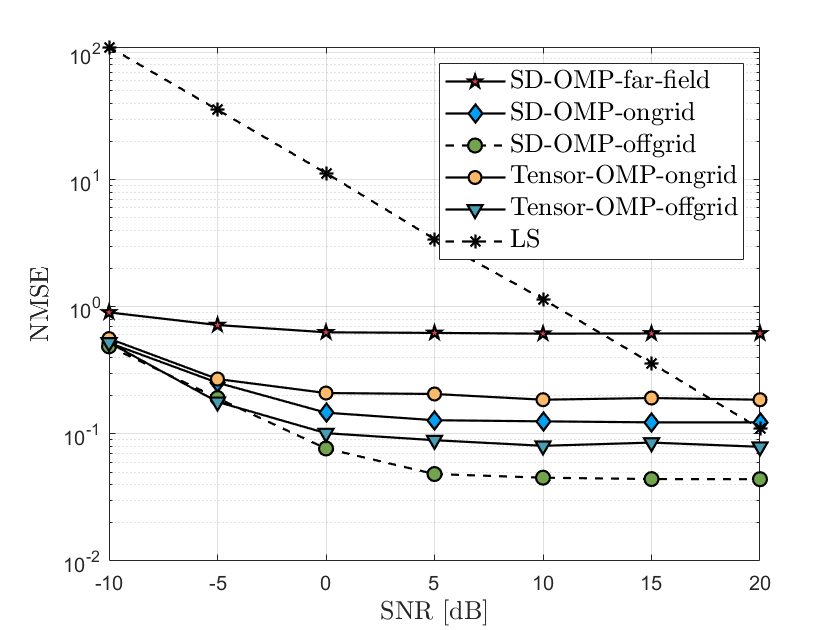}%
\label{fig6_forth_case}}
\caption{NMSE performance under $L=1$ for various methods versus the number of measurements $Q_x/Q_y$. (a) $Q_x=Q_y=6$. (b) $Q_x=Q_y=8$. (c) $Q_x=Q_y=10$. (d) $Q_x=Q_y=12$.}
\label{fig_6}
\vspace{-0.5cm} 
\end{figure*}

\begin{figure*}[!t]
\vspace{-0.35cm} 
\centering
\subfloat[]{\includegraphics[width=0.39\linewidth]{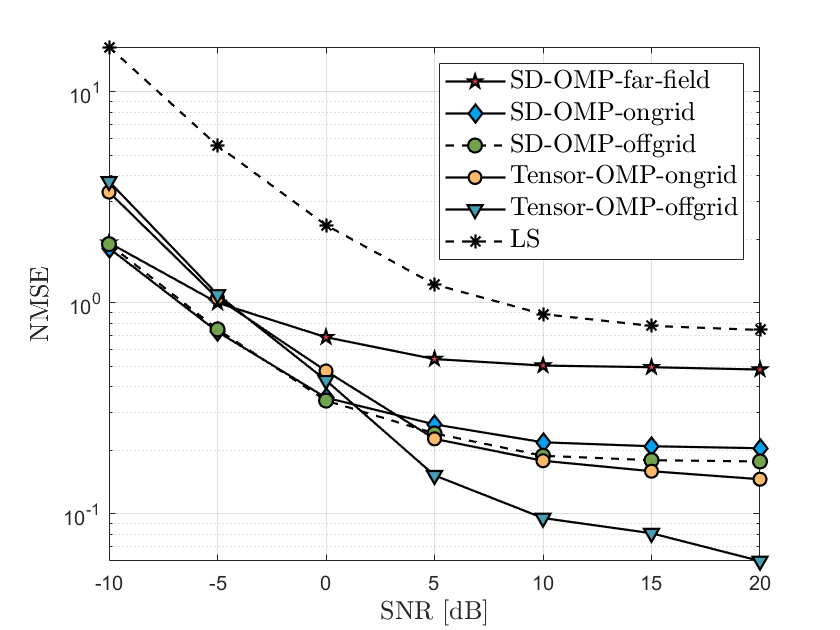}%
\label{fig7_first_case}}
\hfil
\subfloat[]{\includegraphics[width=0.39\linewidth]{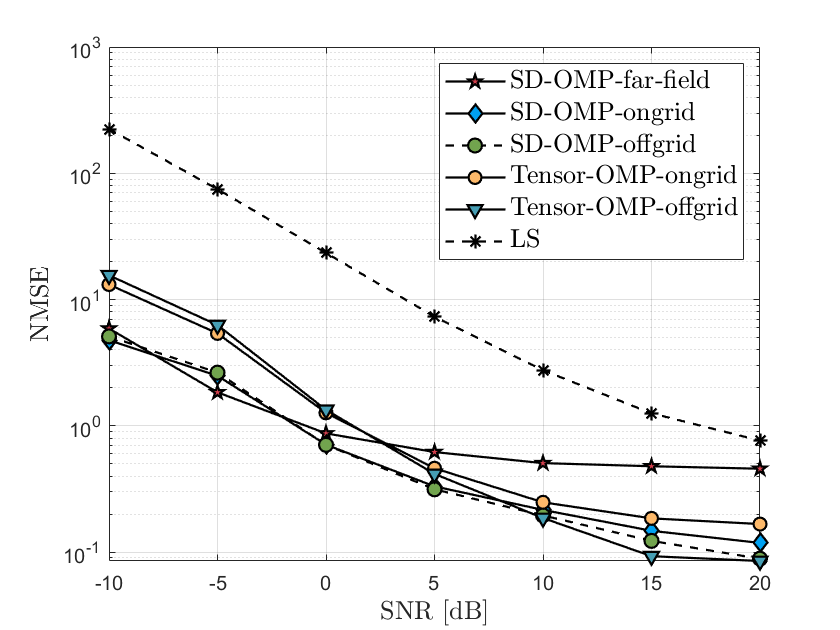}%
\label{fig7_second_case}}
\hfil
\\
\vspace{-0.18in}
\subfloat[]{\includegraphics[width=0.39\linewidth]{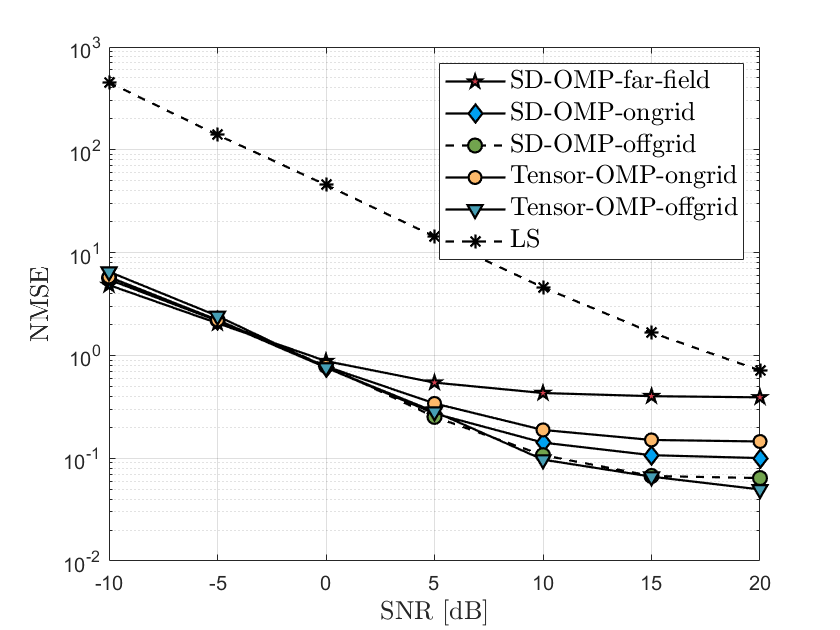}%
\label{fig7_third_case}}
\hfil
\subfloat[]{\includegraphics[width=0.39\linewidth]{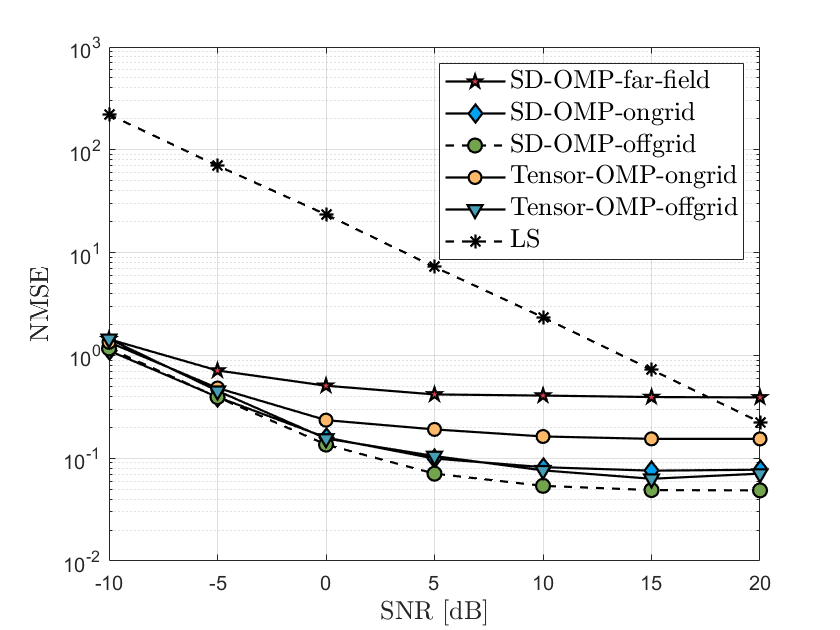}%
\label{fig7_forth_case}}
\caption{NMSE performance under $L=2$ for various methods versus the number of measurements $Q_x/Q_y$. (a) $Q_x=Q_y=6$. (b) $Q_x=Q_y=8$. (c) $Q_x=Q_y=10$. (d) $Q_x=Q_y=12$.}
\label{fig_7}
\vspace{-0.5cm} 
\end{figure*}

\section{ Simulation results}
In this section, several numerical simulations are conducted to demonstrate the effectiveness of the proposed method under various system parameters. The system operates at a central frequency of 10 GHz. The multi-UAV configuration is $M_x=4$, $M_y=2$, with the total number of RF chains being $M=M_x M_y=8$. Each UAV is equipped with UPA of $N_x=N_y=12$. The inter-antenna spacing within the UAV is $d=\lambda/2$, and the spacing between the UAVs in the $x$ and $y$ directions is $D_xd=D_yd=50d$.  Consequently, the system aperture $A_p$ is calculated as $A_p=\sqrt{A_{px}^2 + A_{py}^2}$, where $A_{py}=(M_y-1) D_y d+ (N_y-1) d$ denotes the aperture alone the $y$ axis. So the Rayleigh distance can be computed as $R=2 A_p^2/\lambda=444.63$ m. Furthermore, the azimuth angle $ \cos \theta$, the elevation angle $\sin \varphi $ and the distance $r$ follow the uniform distribution \( \mathcal{U}\left(-0.75, 0.75\right) \), \( \mathcal{U}\left(-0.75, 0.75\right) \) and \( \mathcal{U}\left(50, 80\right) \), respectively. Meanwhile,
when constructing the dictionary, the range of the sampled azimuth angle
$\sin\varphi$ is $[-0.75, 0.75]$, the range of the sampled elevation angle
$\cos\theta$ is $[-0.75, 0.75]$, with the grid resolution set at 
$2/N$, and the range of sampled distance $r$ is [50 m, 80 m]. The distance sampling method in the polar-domain dictionary construction follows the non-uniform strategy proposed in \cite{NF1}, where the sampling is uniformly performed on $1/r$ instead of 
$r$, to reduce the column coherence and improve near-field sparsity, we have 
\begin{small}
\begin{equation}
\setlength{\abovedisplayskip}{-1.2pt}
\boldsymbol{r} = \left[ \frac{1}{1/r_{\max}}, \frac{1}{1/r_{\max} + \Delta}, \ldots, \frac{1}{1/r_{\min}} \right]
\end{equation}
\end{small}
The step size 
 $\Delta$ is set to 0.0003.
\footnote{Distance Sampling:
 The chosen step size satisfies the condition \( \Delta < \frac{1}{Z_\Delta} \) to control dictionary coherence, where $Z_\Delta = D^2/2 \beta^2 \lambda.$
For our system setup with estimated \( Z_\Delta \approx 1235 \)( the desired coherence leve $\beta=0.03$), the maximum allowable step size is \( \Delta_{\text{max}} \approx 0.00081 \), validating our use of \( \Delta = 0.0003 \).}

The benchmark scheme used in the study is described as follows:

\begin{itemize}
    \item \textbf{SD-OMP-far-field}: The algorithm can be understood as the application of the \textbf{SD-OMP-offgrid} in the far field context. $r$ is set as $r=10^5 \text{m} \gg R$. 
    \item \textbf{SD-OMP-ongrid}: As shown in \text{IV}.A.
    \item \textbf{SD-OMP-offgrid}: As shown in IV.A. Integrating the offgrid components in Algorithm 1 based on the SD-OMP-ongrid algorithm. 
    \item \textbf{Tensor-OMP-ongrid}: As shown in Algorithm 1. $\widehat{\boldsymbol{h}}$ is the channel reconstructed by tensor-OMP-ongrid.
    \item \textbf{Tensor-OMP-offgrid}: As shown in Algorithm 1. $\widetilde{\widehat{\boldsymbol{h}}}$ is the channel reconstructed by tensor-OMP-offgrid.
    \item \textbf{LS}: Using the least squares (LS) estimator as an upper bound.
\end{itemize}

As depicted in Fig. 6, we present the normalized mean square error (NMSE) for various methods under conditions where \( L = 1 \), \( Q_x = Q_y \in \{6, 8, 10, 12\} \), and the  \( \text{SNR} \in \{-10, -5, 0, 5, 10, 15, 20\} \) dB. NMSE can be calculated as $\text{NMSE} = \frac{\mathbb{E}(\| \boldsymbol{h} - \hat{\boldsymbol{h}}\|_2^2)}{\mathbb{E}(\| \boldsymbol{h} \|_2^2)}$. It is evident from Fig. 6(a) that at lower SNR levels, both tensor-OMP-ongrid and tensor-OMP-offgrid exhibit slightly higher NMSE compared to SD-OMP-ongrid and SD-OMP-offgrid. However, as the SNR increases, SD-OMP-offgrid and tensor-OMP-offgrid demonstrate superior performance over their ongrid counterparts. Notably, SD-OMP-ongrid shows the poorest performance, while tensor-OMP-offgrid achieves the best result. It is important to observe that with the increase of SNR, tensor-OMP-ongrid approaches a performance similar to that of SD-OMP-ongrid, highlighting the superiority of the tensor-OMP algorithm proposed in this paper. In Fig. 6(b), similar conclusions can be drawn. In Fig. 6(c) and Fig. 6(d), with the increase in the number of measurements \( Q_x = 10, 12 \), SD-OMP-offgrid shows better performance. Although tensor-OMP-ongrid exhibits inferior performance, comparatively, regardless of the variation in \( Q_x \), tensor-OMP-offgrid consistently outperforms tensor-OMP-ongrid in reducing NMSE more significantly than SD-OMP-offgrid does compared to SD-OMP-ongrid. This suggests that tensor-OMP-offgrid is more effective in minimizing NMSE across different values of \( Q \).

In Fig. 7, we plot the NMSE for various methods when $L=2$. At this point, considering the characteristics of UAV communication, channel model (\ref{CM}) comprises one LOS path and one Non-LOS (NLoS) path. As the number of paths $L$ increases, the complexity of the recovery problem also increases. The enhanced sparsity leads to a greater number of parameters that need to be estimated, which in turn causes changes in the NMSE for all methods. In general, Fig. 7 shows similar performance to that of Fig. 6. It can be observed that SD-OMP-ongrid performs better than tensor-OMP-ongrid when $L=2$. On this basis, SD-OMP-offgrid achieves a greater improvement in performance, reflecting the superiority of the SD-OMP algorithm. However, as the SNR increases, the tensor-OMP-offgrid algorithm, in most cases, realizes performance comparable to that of SD-OMP-offgrid, making it suitable for practical applications. Notably, as SNR increases, the NMSE curve gradually flattens, due to the limited number of measurements, which prevents NMSE from approaching zero even as SNR tends to infinity.

\section{Conclusion}
In this paper, we investigated the SNR analysis and channel estimation for multi-UAV near-field systems, which can be regarded as a distributed ELAA architecture. By introducing the HSPWM model, we demonstrated through SNR analysis that it offers a simplified yet accurate framework for near-field channel modeling. For channel estimation, we developed SD-OMP and tensor-OMP approaches. SD-OMP extends traditional PD-OMP from angle-distance to elevation-azimuth-distance estimation, resulting in both ongrid and offgrid variants. Based on HSPWM, we further constructed a tensor dictionary and proposed tensor-OMP-on/offgrid algorithms. Simulation results show that tensor-OMP-offgrid achieves a good trade-off between computational complexity and estimation accuracy. As UAV jitter and synchronization errors are inevitable in practical deployments, understanding and mitigating their impact on system performance will be the focus of our future research.

{\appendices
\section{Proof of (\ref{2int})}
We first define a continuous function $f(x,y,s,t)$ about  $x,y,s,t$ over the 4-directional hypercube, where $x \in \left[-\frac{\xi}{2}, (M_x-\frac{1}{2}) \xi \right]$, $y \in \left[-\frac{\xi}{2}, (M_y-\frac{1}{2}) \xi \right]$, $s \in \left[-\frac{\xi}{2}, (N_x-\frac{1}{2})  \xi \right]$ and $t \in \left[-\frac{\xi}{2}, (N_y-\frac{1}{2}) \xi \right]$, shown at (\ref{func}). Given that $\xi \ll 1$,  within each unit hypercube, where $\forall x \in \left[ \left( m_x - \frac{1}{2} \right) \xi, \left( m_x + \frac{1}{2} \right) \xi \right],  y \in \left[ \left( m_y - \frac{1}{2} \right) \xi, \left( m_y + \frac{1}{2} \right) \xi \right], s \in \left[ \left( n_x - \frac{1}{2} \right) \xi, \left( n_x + \frac{1}{2} \right) \xi \right], t \in \left[ \left( n_y - \frac{1}{2} \right) \xi, \left( n_y + \frac{1}{2} \right) \xi \right]$, we have 
$f(x,y,s,t)\approx f(m_x\xi,m_y\xi,n_x\xi,n_y\xi)$.
\begin{figure*}[ht]
\vspace{-0.2cm}
\normalsize
\centering
\begin{small}
\begin{equation}\label{func}
\setlength{\abovedisplayskip}{-1.5pt}
        f(x,y,s,t)=\frac{1}{1 + 2\Psi D_x x + 2\Psi s + 2\Phi D_y y + 2\Phi t + D_x^2 x^2 + 2D_x xs + s^2 + D_y^2 y^2 + 2D_y yt + t^2}
\end{equation}
\end{small}
\vspace{-0.7cm}
{\rule{\linewidth}{0.4pt}}
\end{figure*}

Based on the definition of Riemann sums, we have 
\begin{small}
\begin{equation}\label{int}
\setlength{\abovedisplayskip}{0.2pt}
\setlength{\belowdisplayskip}{0.2pt}
\begin{aligned}
&\sum_{m_x=0}^{M_x-1} \sum_{m_y=0}^{M_y-1} \sum_{n_x=0}^{N_x-1} \sum_{n_y=0}^{N_y-1}f(m_x\xi,m_y\xi,n_x\xi,n_y\xi)\xi^4 \approx\\[-1mm]&
\int_{-\frac{\xi}{2}}^{(M_x-\frac{1}{2}) \xi}\int_{-\frac{\xi}{2}}^{(M_y-\frac{1}{2}) \xi}\int_{-\frac{\xi}{2}}^{(N_x-\frac{1}{2}) \xi}\int_{-\frac{\xi}{2}}^{(N_y-\frac{1}{2}) \xi} \\[-1mm]&f(x,y,s,t)\, dt \, ds \, dy \, dx\textcolor{blue}{.}
\end{aligned}
\end{equation}
\end{small}

By substituting (\ref{func}) into (\ref{int}), we have (\ref{comfunc}), shown at the top of the page, where (a) holds according to $\int \frac{dx}{A + 2Bx + Cx^2} = \frac{1}{\sqrt{AC - B^2}} \arctan\left(\frac{Cx + B}{\sqrt{Ac - B^2}}\right)$ and (b) holds from the integral formulas $\int \arctan\left(\frac{x}{a}\right) dx = x \arctan\left(\frac{x}{a}\right) - \frac{a}{2} \ln\left(a^2 + x^2\right)$ in \cite{jifen}, respectively. From (\ref{comfunc}) and (\ref{E}), we have (\ref{2int}). The proof of (\ref{2int}) is completed.

\begin{figure*}[ht]
\normalsize
\setlength{\abovedisplayskip}{-0.3pt}
\setlength{\belowdisplayskip}{0.2pt}
\centering
\begin{small}
\begin{equation}\label{comfunc}
\begin{aligned}
&\int_{-\frac{\xi}{2}}^{(M_x-\frac{1}{2}) \xi}\int_{-\frac{\xi}{2}}^{(M_y-\frac{1}{2}) \xi}\int_{-\frac{\xi}{2}}^{(N_x-\frac{1}{2}) \xi}\int_{-\frac{\xi}{2}}^{(N_y-\frac{1}{2}) \xi}\\&\frac{1}{1 + 2\Psi D_x x + 2\Psi s + 2\Phi D_y y + 2\Phi t + D_x^2 x^2 + 2D_x xs + s^2 + D_y^2 y^2 + 2D_y yt + t^2} \, ds \, dt \, dy \, dx \\
& \stackrel{(a)}{=} \int_{-\frac{\xi}{2}}^{(M_x-\frac{1}{2}) \xi}\int_{-\frac{\xi}{2}}^{(M_y-\frac{1}{2}) \xi}\int_{-\frac{\xi}{2}}^{(N_y-\frac{1}{2}) \xi} \frac{arctan{\frac{(N_{x}-\frac{1}{2})\xi+\Psi+D_{x} x}{\sqrt{1 + 2\Phi D_y y + 2\Phi t + D_y^2 y^2 + 2D_y y t + t^2 - \Psi^2}}}}{\sqrt{1 + 2\Phi D_y y + 2\Phi t + D_y^2 y^2 + 2D_y y t + t^2 - \Psi^2} } \, dt \, dy \, dx \\
&-\int_{-\frac{\xi}{2}}^{(M_x-\frac{1}{2}) \xi}\int_{-\frac{\xi}{2}}^{(M_y-\frac{1}{2}) \xi}\int_{-\frac{\xi}{2}}^{(N_y-\frac{1}{2}) \xi}\frac{arctan{\frac{-\frac{1}{2}\xi+\Psi+D_{x} x}{\sqrt{1 + 2\Phi D_y y + 2\Phi t + D_y^2 y^2 + 2D_y y t + t^2 - \Psi^2}}}}{\sqrt{1 + 2\Phi D_y y + 2\Phi t + D_y^2 y^2 + 2D_y y t + t^2 - \Psi^2} }\, dt \, dy \, dx \\
& \stackrel{(b)}{=}
\frac{1}{D_x} \int_{-\frac{\xi}{2}}^{(M_y-\frac{1}{2}) \xi}\int_{-\frac{\xi}{2}}^{(N_y-\frac{1}{2}) \xi} \left[ h \left( \frac{(M_x-\frac{1}{2}) D_x \xi+\Psi+(N_x-\frac{1}{2}) \xi}{ \sqrt{1 + 2\Phi D_y y + 2\Phi t + D_y^2 y^2 + 2D_y y t + t^2 - \Psi^2}} \right)    \right. \\
& \left.
- h \left( \frac{-\frac{1}{2}\xi D_x+\Psi+(N_x-\frac{1}{2}) \xi}{\sqrt{1 + 2\Phi D_y y + 2\Phi t + D_y^2 y^2 + 2D_y y t + t^2 - \Psi^2}} \right) 
  - h \left( \frac{(M_x-\frac{1}{2}) D_x \xi+\Psi-\frac{1}{2}\xi}{\sqrt{1 + 2\Phi D_y y + 2\Phi t + D_y^2 y^2 + 2D_y y t + t^2 - \Psi^2}} \right) \right. \\
& \left. + h \left( \frac{-\frac{1}{2}\xi D_x+\Psi-\frac{1}{2}\xi}{\sqrt{1 + 2\Phi D_y y + 2\Phi t + D_y^2 y^2 + 2D_y y t + t^2 - \Psi^2}} \right) \right] dt dy
\end{aligned}
\end{equation}
\end{small}
{\rule{\linewidth}{0.4pt}}
\vspace{-0.1cm} 
\end{figure*}

\section{Proof of (\ref{H})}
We first give the analytical expression of $\gamma_{\mathbf{SWM,ULA}}$ under the condition that the UAV carries ULA.
\begin{small}
\begin{equation}
\setlength{\abovedisplayskip}{-0.3pt}
\setlength{\belowdisplayskip}{-0.1pt}
\label{G}
\begin{aligned}
\gamma_{\mathbf{SWM,ULA}} &= \frac{\bar{P} \beta_{0}}{r^2} \sum_{m_x=0}^{M_x-1} \sum_{n_x=0}^{N_x-1} \\[-1mm]
& \frac{1}{1 + 2\xi \Psi(m_x D_x + n_x) + (m_x D_x + n_x)^2 \xi^2} 
\end{aligned}
\end{equation}   
\end{small}

Similar with Appendix A, according to the principles of (a) and (b), we have (\ref{rtrt}).

\begin{figure*}[ht]
\vspace{-0.5cm}
\normalsize
\begin{small}
\begin{equation}\label{rtrt}
\setlength{\abovedisplayskip}{-3.3pt}
\setlength{\belowdisplayskip}{0.2pt}
\begin{aligned}
\gamma_{\mathbf{SWM,ULA}} &
 \approx \frac{1}{\xi^2}  \int_{-\frac{\xi}{2} }^{(M_x-\frac{1}{2}) \xi} \int_{-\frac{\xi}{2} }^{(N_x-\frac{1}{2}) \xi} \frac{1}{1 + 2\Psi(x D_x + s) + (x D_x + s)^2} \, dx \, ds \\
\stackrel{(a)}{=} & \frac{1}{\xi^2 \sqrt{1 - \Psi^2}} \int_{-\frac{\xi}{2} }^{(M_x-\frac{1}{2}) \xi} \arctan\left(\frac{N_x \xi+\Psi+D_x x}{\sqrt{1 - \Psi^2}}\right) \, dx 
- \frac{1}{\xi^2 \sqrt{1 - \Psi^2}} \int_{-\frac{\xi}{2} }^{(M_x-\frac{1}{2}) \xi} \arctan\left(\frac{\Psi+D_x x}{\sqrt{1 - \Psi^2}}\right) \, dx
\\
\stackrel{(b)}{=} & \frac{1}{\xi^2 D_x} \left[ h\left(\frac{{(N_x-\frac{1}{2})  \xi+\Psi+D_x (M_x-\frac{1}{2}) \xi}}{ \sqrt{1 - \Psi^2}} \right)  -h\left(\frac{(N_x-\frac{1}{2})  \xi+\Psi-\frac{1}{2}D_x \xi}{\sqrt{1 - \Psi^2}} \right) \right. \\
& \left.
- h\left(\frac{{-\frac{1}{2} \xi+\Psi+(M_x-\frac{1}{2}) D_x\xi}}{\sqrt{1 - \Psi^2}}\right)   +h\left(\frac{{-\frac{1}{2} \xi+\Psi-\frac{1}{2} D_x\xi}}{\sqrt{1 - \Psi^2}}\right) \right] &
\end{aligned}
\end{equation}
\end{small}
\vspace{-0.75cm} 
{\rule{\linewidth}{0.4pt}}
\end{figure*}

\section{Proof of the (\ref{Iii})}
The SNR expression of $\gamma_{\mathbf{HSPW,ULA}}$ under the condition that UAV carries ULA can be given by
\begin{small}
\begin{equation}\label{Fo}
\begin{aligned}
\gamma_{\mathbf{HSPW,ULA}}& = \bar{P} \| \boldsymbol{b}(\theta, \varphi, r) \otimes \boldsymbol{a}(\theta, \varphi) \|^2 \\[-1mm]
&= \frac{\bar{P} \beta_0 N}{r^2 } \sum_{m_x=0}^{M_x-1} \frac{1}{  1 + 2\xi \Psi m_x D_x  +  (m_x D_x)^2  \xi^2 }
\end{aligned}
\vspace{-0.1cm}
\end{equation}
\end{small}
The remaining proof of (\ref{Iii}) is similar to (a) of Appendix A, which is omitted for brevity.
}

 
%

\bibliographystyle{IEEEtran}
\bibliography{main.bib}

\begin{thebibliography}{10}
\providecommand{\url}[1]{#1}
\csname url@samestyle\endcsname
\providecommand{\newblock}{\relax}
\providecommand{\bibinfo}[2]{#2}
\providecommand{\BIBentrySTDinterwordspacing}{\spaceskip=0pt\relax}
\providecommand{\BIBentryALTinterwordstretchfactor}{4}
\providecommand{\BIBentryALTinterwordspacing}{\spaceskip=\fontdimen2\font plus
\BIBentryALTinterwordstretchfactor\fontdimen3\font minus \fontdimen4\font\relax}
\providecommand{\BIBforeignlanguage}[2]{{%
\expandafter\ifx\csname l@#1\endcsname\relax
\typeout{** WARNING: IEEEtran.bst: No hyphenation pattern has been}%
\typeout{** loaded for the language `#1'. Using the pattern for}%
\typeout{** the default language instead.}%
\else
\language=\csname l@#1\endcsname
\fi
#2}}
\providecommand{\BIBdecl}{\relax}
\BIBdecl

\bibitem{6G2}
C.~Wang \emph{et~al.}, ``{On the Road to 6G: Visions, Requirements, Key Technologies, and Testbeds},'' \emph{IEEE Communications Surveys \& Tutorials}, vol.~25, no.~2, pp. 905--974, 2023.

\bibitem{art}
Y.~Xiaohu \emph{et~al.}, ``{Towards 6G wireless communication networks: vision, enabling technologies, and new paradigm shifts},'' \emph{Science China Information Sciences}, vol.~64, 01 2021.

\bibitem{xiao2020overviewintegratedlocalizationcommunication}
\BIBentryALTinterwordspacing
Z.~Xiao and Y.~Zeng, ``{An Overview on Integrated Localization and Communication Towards 6G},'' 2020. [Online]. Available: \url{https://arxiv.org/abs/2006.01535}
\BIBentrySTDinterwordspacing

\bibitem{ISAC}
F.~Liu \emph{et~al.}, ``{Integrated Sensing and Communications: Toward Dual-Functional Wireless Networks for 6G and Beyond},'' \emph{IEEE Journal on Selected Areas in Communications}, vol.~40, no.~6, pp. 1728--1767, 2022.

\bibitem{UR}
S.~Solanki, J.~Park, and I.~Lee, ``{On the Performance of IRS-Aided UAV Networks With NOMA},'' \emph{IEEE Transactions on Vehicular Technology}, vol.~71, no.~8, pp. 9038--9043, 2022.

\bibitem{NF}
H.~Lu \emph{et~al.}, ``{A Tutorial on Near-Field XL-MIMO Communications Toward 6G},'' \emph{IEEE Communications Surveys \& Tutorials}, vol.~26, no.~4, pp. 2213--2257, 2024.

\bibitem{ELAAdai}
M.~Cui and L.~Dai, ``{Near-Field Wideband Beamforming for Extremely Large Antenna Arrays},'' \emph{IEEE Transactions on Wireless Communications}, vol.~23, no.~10, pp. 13\,110--13\,124, 2024.

\bibitem{XLarray}
Y.~Zhang, C.~You, L.~Chen, and B.~Zheng, ``{Mixed Near- and Far-Field Communications for Extremely Large-Scale Array: An Interference Perspective},'' \emph{IEEE Communications Letters}, vol.~27, no.~9, pp. 2496--2500, 2023.

\bibitem{PWM1}
W.~Wang and W.~Zhang, ``{Jittering Effects Analysis and Beam Training Design for UAV Millimeter Wave Communications},'' \emph{IEEE Transactions on Wireless Communications}, vol.~21, no.~5, pp. 3131--3146, 2022.

\bibitem{1}
W.~Chen, C.~Liu, W.~Wang, M.~Peng, and W.~Zhang, ``{Adaptive Hybrid Beamforming for UAV mmWave Communications Against Asymmetric Jitter},'' \emph{IEEE Transactions on Wireless Communications}, vol.~23, no.~8, pp. 9432--9445, 2024.

\bibitem{swarm}
B.~Shang, E.~S. Bentley, and L.~Liu, ``{UAV Swarm-Enabled Aerial Reconfigurable Intelligent Surface: Modeling, Analysis, and Optimization},'' \emph{IEEE Transactions on Communications}, vol.~71, no.~6, pp. 3621--3636, 2023.

\bibitem{swarmmm}
N.~Deng \emph{et~al.}, ``{Enhancing Millimeter Wave Cellular Networks via UAV-Borne Aerial IRS Swarms},'' \emph{IEEE Transactions on Communications}, vol.~72, no.~1, pp. 524--538, 2024.

\bibitem{ELAAtutorial}
Z.~Wang \emph{et~al.}, ``{A Tutorial on Extremely Large-Scale MIMO for 6G: Fundamentals, Signal Processing, and Applications},'' \emph{IEEE Communications Surveys \& Tutorials}, vol.~26, no.~3, pp. 1560--1605, 2024.

\bibitem{HOWDOES}
H.~Lu and Y.~Zeng, ``{How Does Performance Scale with Antenna Number for Extremely Large-Scale MIMO?}'' in \emph{ICC 2021 - IEEE International Conference on Communications}, 2021, pp. 1--6.

\bibitem{elaaa}
H.~\vspace{0mm}Lu and Y.~Zeng, ``{Communicating With Extremely Large-Scale Array/Surface: Unified Modeling and Performance Analysis},'' \emph{IEEE Transactions on Wireless Communications}, vol.~21, no.~6, pp. 4039--4053, 2022.

\bibitem{yang2}
S.~Yang \emph{et~al.}, ``{Near-Field Channel Estimation for Extremely Large-Scale Reconfigurable Intelligent Surface (XL-RIS)-Aided Wideband mmWave Systems},'' \emph{IEEE Journal on Selected Areas in Communications}, vol.~42, no.~6, pp. 1567--1582, 2024.

\bibitem{yang3}
S.~Yang, W.~Lyu, Z.~Hu, Z.~Zhang, and C.~Yuen, ``{Channel Estimation for Near-Field XL-RIS-Aided mmWave Hybrid Beamforming Architectures},'' \emph{IEEE Transactions on Vehicular Technology}, vol.~72, no.~8, pp. {11\,029--11\,034}, 2023.

\bibitem{modular}
X.~Li, H.~Lu, Y.~Zeng, S.~Jin, and R.~Zhang, ``{Near-Field Modeling and Performance Analysis of Modular Extremely Large-Scale Array Communications},'' \emph{IEEE Communications Letters}, vol.~26, no.~7, pp. 1529--1533, 2022.

\bibitem{WSMS}
S.~Yang \emph{et~al.}, ``{Performance Bounds for Near-Field Localization With Widely-Spaced Multi-Subarray mmWave/THz MIMO},'' \emph{IEEE Transactions on Wireless Communications}, vol.~23, no.~9, pp. 10\,757--10\,772, 2024.

\bibitem{WSMSS}
L.~Yan, Y.~Chen, C.~Han, and J.~Yuan, ``{Joint Inter-Path and Intra-Path Multiplexing for Terahertz Widely-Spaced Multi-Subarray Hybrid Beamforming Systems},'' \emph{IEEE Transactions on Communications}, vol.~70, no.~2, pp. 1391--1406, 2022.

\bibitem{WS}
Y.~Chen, L.~Yan, and C.~Han, ``{Hybrid Spherical- and Planar-Wave Modeling and DCNN-Powered Estimation of Terahertz Ultra-Massive MIMO Channels},'' \emph{IEEE Transactions on Communications}, vol.~69, no.~10, pp. 7063--7076, 2021.

\bibitem{WSS}
Y.~Chen, R.~Li, C.~Han, S.~Sun, and M.~Tao, ``{Hybrid Spherical- and Planar-Wave Channel Modeling and Estimation for Terahertz Integrated UM-MIMO and IRS Systems},'' \emph{IEEE Transactions on Wireless Communications}, vol.~22, no.~12, pp. 9746--9761, 2023.

\bibitem{crossfield}
J.~Luo, J.~Fan, K.~Xie, and X.~Shi, ``Efficient hybrid near- and far-field beam training for xl-mimo communications,'' \emph{IEEE Transactions on Vehicular Technology}, vol.~73, no.~12, pp. 19\,785--19\,790, 2024.

\bibitem{yang5}
S.~Yang, H.~Chen, W.~Liu, X.-P. Zhang, and C.~Yuen, ``Near-field channel estimation and localization: Recent developments, cooperative integration, and future directions,'' \emph{IEEE Signal Processing Magazine}, vol.~42, no.~1, pp. 60--73, 2025.

\bibitem{NF1}
M.~Cui and L.~Dai, ``{Channel Estimation for Extremely Large-Scale MIMO: Far-Field or Near-Field?}'' \emph{IEEE Transactions on Communications}, vol.~70, no.~4, pp. 2663--2677, 2022.

\bibitem{NF3}
X.~Guo, Y.~Chen, and Y.~Wang, ``{Compressed Channel Estimation for Near-Field XL-MIMO Using Triple Parametric Decomposition},'' \emph{IEEE Transactions on Vehicular Technology}, vol.~72, no.~11, pp. 15\,040--15\,045, 2023.

\bibitem{yang1}
S.~Yang \emph{et~al.}, ``{Near-field channel estimation for extremely large-scale Terahertz communications},'' \emph{SCIENCE CHINA Information Sciences}, vol.~67, no.~9, pp. 192\,302--, 2024.

\bibitem{ZhangTMC2025}
C.~Zhang \emph{et~al.}, ``{Multi-Objective Aerial Collaborative Secure Communication Optimization via Generative Diffusion Model-Enabled Deep Reinforcement Learning},'' \emph{IEEE Transactions on Mobile Computing}, vol.~24, no.~4, pp. 3041--3058, 2025.

\bibitem{LiuTMC2024}
S.~Liu \emph{et~al.}, ``{UAV-Enabled Collaborative Beamforming via Multi-Agent Deep Reinforcement Learning},'' \emph{IEEE Transactions on Mobile Computing}, vol.~23, no.~12, pp. 13\,015--13\,032, 2024.

\bibitem{tongbu}
S.~Jayaprakasam, S.~K.~A. Rahim, and C.~Y. Leow, ``Distributed and collaborative beamforming in wireless sensor networks: Classifications, trends, and research directions,'' \emph{IEEE Communications Surveys \& Tutorials}, vol.~19, no.~4, pp. 2092--2116, 2017.

\bibitem{ref2}
R.~G. Stephen and R.~Zhang, ``Uplink channel estimation and data transmission in millimeter-wave cran with lens antenna arrays,'' \emph{IEEE Transactions on Communications}, vol.~66, no.~12, pp. 6542--6555, 2018.

\bibitem{PWM2}
K.~Xu \emph{et~al.}, ``{Channel Feature Projection Clustering Based Joint Channel and DoA Estimation for ISAC Massive MIMO OFDM System},'' \emph{IEEE Transactions on Vehicular Technology}, vol.~73, no.~3, pp. 3678--3689, 2024.

\bibitem{pwm3}
K.~Kim and J.~P. Choi, ``{3D Network Design for Multi-UAV RAN With THz-Empowered Backhaul},'' \emph{IEEE Transactions on Wireless Communications}, vol.~23, no.~11, pp. 16\,437--16\,452, 2024.

\bibitem{Xiong}
B.~Xiong, Z.~Zhang, C.~Pan, and J.~Wang, ``{Performance Analysis of Aerial RIS Auxiliary mmWave Mobile Communications With UAV Fluctuation},'' \emph{IEEE Wireless Communications Letters}, vol.~13, no.~4, pp. 1183--1187, 2024.

\bibitem{NMdanchuan}
J.~C. Lagarias, J.~A. Reeds, M.~H. Wright, and P.~E. Wright, ``{Convergence Properties of the Nelder--Mead Simplex Method in Low Dimensions},'' \emph{SIAM Journal on Optimization}, vol.~9, no.~1, pp. 112--147, 1998.

\bibitem{3DOMP}
H.~Yingqiu, F.~Yong, and H.~Lei, ``{3D sparse signal recovery via 3D orthogonal matching pursuit},'' \emph{Journal of Systems Architecture}, vol.~64, pp. 3--10, 2016, real-Time Signal Processing in Embedded Systems.

\bibitem{SDAR}
Y.-T. Chiew and Y.-P. Lin, ``{Channel Estimation for Hybrid mmWave Systems Using Generalized Kronecker Compressive Sensing (G-KCS) With Successive Decision-Aided Recovery},'' \emph{IEEE Transactions on Signal Processing}, vol.~72, pp. 2970--2982, 2024.

\bibitem{2DOMP}
Y.~Fang, J.~Wu, and B.~Huang, ``{2D sparse signal recovery via 2D orthogonal matching pursuit},'' \emph{Science China Information Sciences}, vol.~55, pp. 889--897, 2012.

\bibitem{jifen}
I.~S. Gradshteyn and I.~M. Ryzhik, \emph{Table of integrals, series, and products}.\hskip 1em plus 0.5em minus 0.4em\relax Academic press, 2014.

\end{thebibliography}

\vfill

\end{document}